\documentclass[11pt]{article}
\usepackage{times,graphicx,theorem,epsfig,lscape,dcolumn,multirow,epic,float}
\usepackage{amsmath,amssymb}
\usepackage[normalem]{ulem}
\usepackage{subfigure}
\usepackage{adjustbox}
\usepackage{natbib}
\usepackage[figuresright]{rotating}
\usepackage[english]{babel}
\usepackage{booktabs}
\usepackage{hhline}
\usepackage{bm}
\usepackage{enumerate}
\usepackage{verbatim}
\usepackage{dsfont}
\usepackage[colorlinks=true,linkcolor=black,citecolor=black]{hyperref}
\usepackage{threeparttable}
\usepackage{array}
\usepackage{comment}
\usepackage{pdflscape}
\usepackage{afterpage}
\usepackage{booktabs}
\usepackage{moreverb}
\usepackage{graphicx}
\usepackage{tikz}
\usepackage[labelsep=quad,skip=-2pt]{caption}

\numberwithin{equation}{section}
\theoremstyle{plain} 
\theoremstyle{plain} 
\theoremstyle{plain} \newtheorem{remark}{{\sc Remark}}
\theoremstyle{plain} 
\theoremstyle{plain} 

\theoremstyle{plain}

\theoremstyle{plain} 

\theoremstyle{plain}

\newcommand{\blinding}[2]{#1}   

\newcommand{\mX}{\mathcal{X}}

\newcommand{\mR}{\mathcal{R}}

\newcommand{\mY}{\mathcal{Y}}

\newcommand{\mM}{\mathcal{M}}
\newcommand{\mO}{\mathcal{O}}

\theoremstyle{plain} \newtheorem{theory}{{\sc Theory}}

\usepackage{tikz-cd}
\tikzset{smalltext/.style={"\textup{\small #1}" description}}

\begin{document}

\begin{center}
\vspace*{-2.5cm}
{\Large Covariate-adjusted win statistics in randomized clinical trials with ordinal outcomes}

\medskip
\blinding{
Zhiqiang Cao$^{1}$, Scott Zuo$^{2}$, Mary Ryan Baumann$^{3,4}$, Kendra Plourde$^{5,6}$, Patrick Heagerty$^{7,8}$, \\ Guangyu Tong$^{5,9,10,11}$ and Fan Li$^{5,9,10,11,*}$
}{}

$^{1}$School of Artificial Intelligence, Shenzhen Technology University, Guangdong, P. R. China \\
$^{2}$Department of Preventive Medicine, Feinberg School of Medicine, Northwestern University, Illinois, U.S.A \\
$^{3}$Department of Biostatistics and Medical Informatics, University of Wisconsin - Madison, Wisconsin, U.S.A\\
$^{4}$Department of Population Health Sciences, University of Wisconsin - Madison, Wisconsin, U.S.A\\
$^{5}$Department of Biostatistics, Yale University School of Public Health, Connecticut, U.S.A\\
$^{6}$Yale Center for Analytical Sciences, Yale University School of Public Health, Connecticut, U.S.A\\
$^{7}$Department of Biostatistics, University of Washington, Washington, U.S.A\\
$^{8}$Clinical Learning, Evidence, and Research (CLEAR) Center, University of Washington, Washington, U.S.A\\
$^{9}$Section of Cardiovascular Medicine, Department of Internal Medicine, Yale School of Medicine, Connecticut, U.S.A\\
$^{10}$Cardiovascular Medicine Analytics Center, Yale School of Medicine, Connecticut, U.S.A\\	
$^{11}$Center for Methods in Implementation and Prevention Science, Yale School of Public Health, Connecticut, U.S.A\\
$*$fan.f.li@yale.edu\\

\end{center}

\date{}

{\centerline{Abstract}
\noindent Ordinal outcomes are common in clinical settings where they often represent increasing levels of disease progression or different levels of functional impairment. In this article, we focus on representing the average treatment effect for ordinal outcomes via intrinsic pairwise outcome comparisons captured through win estimands, such as the win ratio and win difference. Recognizing the value of baseline covariate adjustment toward enhanced precision, we first develop propensity score weighting estimators, including both inverse probability weighting (IPW) and overlap weighting (OW), tailored to estimating win estimands. Furthermore, we develop augmented weighting estimators that leverage an additional ordinal outcome regression to potentially improve efficiency over weighting alone. Leveraging the theory of U-statistics, we establish the asymptotic theory for all estimators, and derive closed-form variance estimators to support statistical inference. We also prove that all of the covariate-adjusted estimators do not compromise consistency for the target estimand even when the associated working models are incorrectly specified; hence these covariate-adjusted estimators are model-robust. Through simulations we demonstrate the enhanced efficiency of the weighted estimators over the unadjusted estimator, with the augmented weighting estimators showing a further improvement in efficiency except for extreme cases. Finally, we illustrate our proposed methods with the ORCHID trial, and implement our covariate adjustment methods in an R package \protect\texttt{winPSW}.
\vspace*{0.3cm}

}

\clearpage

\section{Introduction}\label{s:intro}
Ordinal outcomes are common in randomized clinical trials (RCTs) and are often comprised of multiple, monotonically ordered categories that do not necessarily align with a quantifiable distance \citep{Selman2023review}. For example, the FLU-IVIG trial \citep{Peterson2017ordinal} considered an ordinal outcome constructed with six mutually exclusive categories: (1) death; (2) intensive care unit (ICU) hospitalization (in ICU); (3) non-ICU hospitalization, requiring supplemental oxygen; (4) non-ICU hospitalization, not requiring supplemental oxygen; (5) discharged from the hospital but unable to resume normal activities; and (6) discharged from the hospital with resumption of normal activities. The categories were defined to delineate clear
improvement and worsening in patient status and to
yield a sufficient spread of the data for showing benefit
due to intravenous hyperimmune immunoglobulin. Ordinal outcomes are useful in clinical settings, as specific disease states can represent meaningfully distinct categories that may be of clinical importance to researchers. However, the statistical considerations for analyzing ordinal outcomes can often be more complicated, especially when it comes to defining an assumption-lean target parameter to quantify the treatment effect. Conventional methods are often based on parametric ordinal regression models (e.g., proportional odds model) with an odds ratio target parameter \citep{Uddin2024PO}; however, if the proportional odds assumption fails to hold, the estimated odds ratio may be
subject to bias with ambiguous interpretation. Instead, 
alternative summary estimands/measures such as the win ratio \citep{Pocock2012,Mao2018,Klok2023}, win odds (WO) \citep{Dong2020wo} --- a modification of WR
	to include ties, and win difference (WD, or net benefit) \citep{buyse2010generalized} have gained traction in recent years. \cite{Evans2020} showed that the win ratio can more optimally
inform patient treatment, enhance the understanding of the totality of intervention effects on patients, and potentially
provide efficiencies over standard analyses in a randomized clinical trial. 

In this article, we focus on win statistics-based analyses of ordinal outcomes in randomized trials, and study covariate adjustment strategies to improve statistical efficiency. 
In RCTs, pre-specified covariate adjustment is a critical strategy for improving the precision of the treatment effect estimator 
as well as a crucial step in trials using randomization strata. Regulatory agencies such as the Food and Drug Administration (FDA) \citep{food148910adjusting} and the European Medicines Agency (EMA) \citep{ICH_E9} have explicitly recommended incorporating baseline covariates to enhance statistical efficiency in treatment comparisons. A rich literature has examined covariate adjustment methods for estimating the average treatment effects in randomized trials. For example, \citet{Benkeser2021covadj} demonstrated that adjusting for baseline covariates in the analysis of binary, ordinal, and time-to-event outcomes can markedly reduce the required sample size by improving statistical precision. \citet{shen2014inverse} and \citet{williams2014variance} showed that inverse probability weighting (IPW) estimators, when applied with prognostic covariates to estimate known treatment propensities under randomization, can reduce the asymptotic variance compared to the unadjusted treatment effect estimator. \citet{zeng2021propensity} proposed overlap weighting (OW) for covariate adjustment in randomized trials, showing that OW estimators are asymptotically equivalent to IPW and ANCOVA estimators under 1:1 randomization but can yield lower finite-sample variance than IPW. \citet{wang2023model} established the consistency and asymptotic normality of a general class of M-estimators that integrate covariate adjustment and stratified randomization, showing substantial gains in efficiency and power. In contrast, far less attention has been devoted to covariate adjustment for pairwise comparison estimands, such as the win ratio. Notable among the sparse contributions, \citet{Vermeulen2015increasing} extended the Mann-Whitney U test using semiparametric efficiency theory to exploit covariate information, leading to improved power and enabling exact permutation-based inference. \citet{zhang2019estimate} investigated several estimators---including regression-based, IPW, and augmented inverse probability weighting (AIPW) approaches---for Mann-Whitney-type causal effects in both clinical trials and observational studies. More recently, \citet{wang2025adjustedWR} proposed inverse probability of treatment weighting (IPTW)-adjusted win ratio estimators for composite time-to-event outcomes, demonstrating improved power under non-proportional hazards. \cite{buyse2025} briefly discussed covariate adjustment methods including IPW and AIPW for generalized pairwise comparison estimands in observational studies. 
While these previous efforts illustrate the potential of covariate adjustment (e.g., IPW) in pairwise comparisons, there is a lack of formal theoretical justification for the asymptotic efficiency gain of IPW in this context. Moreover, a more systematic investigation of a broader class of propensity score weighting or augmented weighting estimators for covariate-adjusted win analysis, particularly for ordinal outcomes, remains unavailable in the existing literature.

To address this gap, we study a general class of propensity score weighting methods for baseline covariate adjustment when using win statistics to analyze RCTs with ordinal outcomes. The key contributions of this work are threefold. First, we propose a class of covariate-adjusted weighting estimators for win statistics based on pairwise comparisons, incorporating both IPW and OW as two special cases, thereby extending the framework of \citet{zeng2021propensity} to a broader class of estimands. Second, we discuss augmented weighting estimators that combine any propensity score balancing weights with an outcome model to improve efficiency. This generalizes the AIPW formulation of \citet{Mao2019} to a wider range of balancing weighting schemes. Third, we establish the asymptotic properties of these estimators under randomized assignment by leveraging the theory of M-estimation and U-statistics. We further derive a set of closed-form variance estimators to enable computationally efficient inference without the need for resampling. Through this effort, a key theoretical insight is that the influence function of a balancing weighting estimator can be expressed as the influence function of the unadjusted estimator minus its projection onto the nuisance tangent space, for estimating the pairwise comparison estimands. This decomposition allows us to formally characterize the asymptotic variance reduction achieved through covariate adjustment for the win difference and win ratio, and to justify the use of propensity score weighting when applying win statistics in randomized trials. Through extensive simulations, we demonstrate that the proposed covariate-adjusted estimators, particularly augmented weighting estimators under either IPW or OW, substantially improve the efficiency of the unadjusted estimators for win estimands in randomized trials. We also implement the proposed covariate-adjusted win statistics in an R package \texttt{winPSW} to facilitate practical implementation.

The rest of this article is organized as follows. In Section \ref{sec:unadjusted}, we introduce notation and estimands. In Section \ref{sec:weighting}, we develop covariate-adjusted IPW and OW win estimators and derive their asymptotic properties. We study augmented weighting estimators that leverage an additional outcome regression to further improve efficiency in Section \ref{sec:augmented}. Simulation studies are carried out to examine the finite-sample performance of the proposed estimators relative to the unadjusted estimator in Section \ref{sec:simulation}. Finally, we illustrate our methods by reanalyzing the ordinal outcome data from the Outcomes Related to COVID-19 Treated With Hydroxychloroquine Among Inpatients With Symptomatic Disease (ORCHID) trial in Section \ref{sec:ORCHID}, followed by a discussion in Section \ref{sec:Discussion}.

\section{Win estimands and unadjusted estimators}\label{sec:unadjusted}
\subsection{Notation and estimands}\label{subsec:notations}
We consider a two-arm RCT with a total of $n$ patients. Let $Z$ be the binary treatment assignment, where $Z=1$ for treatment and $Z=0$ for control. Let $n_1=\sum_{i=1}^nZ_i$ and $n_0=n-n_1$, denote the sample size in each arm. Suppose the outcome $Y$ is ordinal with sample space $\mY=\{1,2,\dots,L\}$, where $L$ is a positive finite integer. We assume that larger values of $Y$ are more favorable. Under the potential outcomes framework, each individual has a pair of potential outcomes $\{Y_i(1),Y_i(0)\}$, and we denote the observed outcome as $Y_i =Z_iY_i(1)+(1-Z_i)Y_i(0)$. Let $X=(X_1,\dots,X_p)^{\top}$ be a $p$-dimensional vector of baseline covariates collected for each patient. The observed data then consists of the tuple $\mO_i=(Y_i,Z_i,X_i),~i=1,\dots,n$, which are $n$ independent and identically distributed realizations of $(Y,Z,X)$. With a comparison operator ``$>$'' or ``$<$'', the win estimand is given by \cite{wang2016win}:
\begin{eqnarray}\label{win_prob}
	\tau_{1}=\mathbb{P}\left\{Y_i(1)>Y_j(0)\right\},
\end{eqnarray}
The above estimand has also been referred to by alternative terminology, including the Mann-Whitney effect, relative effect, and desirability of outcome ranking (DOOR) probability \citep{buyse2025}; here we refer it as the win probability --- that is, the probability that a randomly selected $Y_i(1)$ in the study population achieves a more favorable outcome than a randomly selected $Y_j(0)$. If there are no ties or ties are split evenly in both win and loss groups, a value greater than $1/2$ indicates that the treatment leads to more favorable outcomes stochastically. We can similarly define the loss estimand $\tau_{-1}=\mathbb{P}\left\{Y_i(1)<Y_j(0)\right\}$ and write the win ratio (WR) and win difference (WD, or net benefit) estimands, respectively, as:
\begin{eqnarray}\label{win_stat}
	\theta = \frac{\tau_1}{\tau_{-1}}= \frac{\mathbb{P}\left\{Y_i(1)>Y_j(0)\right\}}{\mathbb{P}\left\{Y_i(1)<Y_j(0)\right\}}, \quad    \eta=\tau_1-\tau_{-1}=\mathbb{P}\left\{Y_i(1)>Y_j(0)\right\}-\mathbb{P}\left\{Y_i(1)<Y_j(0)\right\},
\end{eqnarray}
In what follows, we primarily discuss the estimation and inference of the win estimand and the associated covariate adjustment methods, before returning to the WR and WD estimands.

\subsection{Unadjusted estimator}\label{subsec:unadjust}
In a two-sample setting, \citet{Bebu2016} proposed an unbiased covariate-unadjusted estimator for the win estimand $\tau_1$ and loss estimand $\tau_{-1}$, and derived the large-sample results using multivariate multi-sample U-statistic. Below, we review and adapt their results under the potential outcomes framework. 
Under randomization, the treatment assignment $Z_i$ follows a Bernoulli random variable with probability $\pi=\mathbb{P}\left[Z_i=1|X_i,Y_i(1),Y_i(0)\right]=\mathbb{P}(Z_i=1) \in (0,1)$. Under 1:1 randomization, $\pi=1/2$. 
Then the unadjusted win probability estimator is:
\begin{eqnarray}\label{unadjust0}
	\widehat{\tau}_{1}^{\text{UNADJ}} = \frac{\sum_{i=1}^n\sum_{j=1}^nZ_i(1-Z_j)\mathbb{I}(Y_i > Y_j)}{\sum_{i=1}^n\sum_{j=1}^nZ_i(1-Z_j)}.
\end{eqnarray}
Dividing both the numerator and denominator of \eqref{unadjust0} by $n^2$, we have 
$\frac{1}{n^2}\sum_{i=1}^n\sum_{j=1}^nZ_i(1-Z_j)\mathbb{I}(Y_i > Y_j) \stackrel{p}{\longrightarrow}\mathbb{E}\left[Z_i(1-Z_j)\mathbb{I}\left(Y_i(1) > Y_j(0)\right)\right]=\mathbb{E}\left[Z_i(1-Z_j)\right]\mathbb{P}\left\{Y_i(1)>Y_j(0)\right\}$, and 
$\frac{1}{n^2}\sum_{i=1}^n\sum_{j=1}^nZ_i(1-Z_j) \stackrel{p}{\longrightarrow} \mathbb{E}\left[Z_i(1-Z_j)\right]$, where $\mathbb{I}(\cdot)$ is an indicator function; then it is easy to show that $\widehat{\tau}_{1}^{\text{UNADJ}}$ is a consistent estimator of $\tau_{1}$. 

We note that the index $i$ and $j$ in \eqref{unadjust0} are defined relative to the whole randomized sample. In practice, individual $i$ or $j$ can only belong to one of the treatment and control groups. 
If we treat $Y_i(1)$ and $Y_j(0)$ as coming from two groups with sample sizes equal to $n_1$ and $n_0$, respectively (similar to the setup in \citet{Bebu2016}), then $\widehat{\tau}_{1}^{\text{UNADJ}}$ is a multivariate multi-sample U-statistics:
\begin{eqnarray}\label{unadjust}
	\widehat{\tau}_{1}^{\text{UNADJ}} = \frac{1}{n_1n_0} \sum_{i=1}^{n_1} \sum_{j=1}^{n_0} \frac{Z_i(1-Z_j)\mathbb{I}(Y_i > Y_j)}{{n_1}^{-1} \sum_{i=1}^{n_1} Z_i \times {n_0}^{-1} \sum_{j=1}^{n_0} (1-Z_j)}= \frac{1}{n_1n_0} \sum_{i=1}^{n_1} \sum_{j=1}^{n_0} \omega_{ij} \mathbb{I}(Y_i > Y_j),
\end{eqnarray}
where the weight is given by $\omega_{ij} = \frac{Z_i(1-Z_j)}{{n_1}^{-1} \sum_{i=1}^{n_1} Z_i \times {n_0}^{-1} \sum_{j=1}^{n_0} (1-Z_j)}$ and \eqref{unadjust} can be regarded as the weighted form of $U_1$ in (2.2) of \citet{Bebu2016}.  
Then, according to \citet{Bebu2016}, we have
\begin{eqnarray}\label{unadjusted_var}
	\sqrt{n}\left(\widehat{\tau}_{1}^{\text{UNADJ}} - \tau_{1}\right) \stackrel{d}\sim N\left(0, \sigma_{1}^2\right),
\end{eqnarray}
where $\sigma_{1}^2 = \frac{n}{n_1} \xi_{10} + \frac{n}{n_0} \xi_{01}$, $n = n_1 + n_0$ and
\begin{eqnarray*}
	\xi_{10} &=& \text{cov}\left\{\omega_{ij}\mathbb{I}(Y_i > Y_j), \omega_{ik}\mathbb{I}(Y_i > Y_k)\right\}
	= \mathbb{P}\left[Y_i(1) > Y_j(0) ~\&~ Y_i(1) > Y_k(0)\right] - \left\{\mathbb{P}\left[Y_i(1) > Y_j(0)\right]\right\}^2, \\
	\xi_{01} &=& \text{cov}\left\{\omega_{ij}\mathbb{I}(Y_i > Y_j), \omega_{kj}\mathbb{I}(Y_k > Y_j)\right\}
	= \mathbb{P}\left[Y_i(1) > Y_j(0) ~\&~ Y_k(1) >Y_j(0)\right] - \left\{\mathbb{P}\left[Y_i(1) > Y_j(0)\right]\right\}^2.
\end{eqnarray*}

Furthermore, by replacing ``$>$'' with ``$<$'' in \eqref{unadjust}, we obtain the corresponding unadjusted estimator for the loss probability estimand $\tau_{-1}$, denoted as $\widehat{\tau}_{-1}^{\text{UNADJ}}$, with similar large sample properties. Finally, the asymptotic covariance between $\widehat{\tau}_{1}^{\text{UNADJ}}$ and $\widehat{\tau}_{-1}^{\text{UNADJ}}$ is given by $\sigma_{12}=\frac{n}{n_1}\xi^{12}_{10}+\frac{n}{n_0}\xi^{12}_{01}$, where
\begin{eqnarray*}
	\xi^{12}_{10}&=&{\text{cov}}\left\{\omega_{ij}\mathbb{I}(Y_i>Y_j),\omega_{ik}\mathbb{I}(Y_i<Y_k)\right\}\\
	&=&\mathbb{P}\left[Y_i(1)>Y_j(0)~\&~Y_i(1)<Y_k(0)\right]-\left\{\mathbb{P}\left[Y_i(1)>Y_j(0)\right]\right\}\left\{\mathbb{P}\left[Y_i(1)<Y_j(0)\right]\right\}, \\
	\xi^{12}_{01}&=&{\text{cov}}\left\{\omega_{ij}\mathbb{I}(Y_i>Y_j),\omega_{kj}\mathbb{I}(Y_k<Y_j)\right\}\\
	&=&\mathbb{P}\left[Y_i(1)>Y_j(0)~\&~Y_k(1)<Y_j(0)\right]-\left\{\mathbb{P}\left[Y_i(1)>Y_j(0)\right]\right\}\left\{\mathbb{P}\left[Y_i(1)<Y_j(0)\right]\right\}.
\end{eqnarray*}
Once $\sigma_{12}$ is estimated, the covariance between $\widehat{\tau}_{1}^{\text{UNADJ}}$ and $\widehat{\tau}_{-1}^{\text{UNADJ}}$ can be computed as ${\text{cov}}\left(\widehat{\tau}_{1}^{\text{UNADJ}},\widehat{\tau}_{-1}^{\text{UNADJ}}\right)=\frac{1}{n}\widehat{\sigma}_{12}$.
By applying a Taylor expansion, the variance of the estimated WR, $\widehat{\theta}$, based on unadjusted estimator \eqref{unadjust} can be approximated as
\begin{eqnarray}\label{variance_wr}
	{\text{var}}(\widehat{\theta})={\text{var}}(\widehat{\tau}_{1}^{\text{UNADJ}}/\widehat{\tau}_{-1}^{\text{UNADJ}})\approx \frac{(\widehat{\tau}_{1}^{\text{UNADJ}})^2}{(\widehat{\tau}_{-1}^{\text{UNADJ}})^2}\bigg[\frac{{\text{var}}(\widehat{\tau}_{1}^{\text{UNADJ}})}{(\widehat{\tau}_{1}^{\text{UNADJ}})^2}-2\frac{{\text{cov}}(\widehat{\tau}_{1}^{\text{UNADJ}},\widehat{\tau}_{-1}^{\text{UNADJ}})}{\widehat{\tau}_{1}^{\text{UNADJ}}\cdot\widehat{\tau}_{-1}^{\text{UNADJ}}}+\frac{{\text{var}}(\widehat{\tau}_{-1}^{\text{UNADJ}})}{(\widehat{\tau}_{-1}^{\text{UNADJ}})^2}\bigg].
\end{eqnarray}
Note that the variance of the corresponding estimated WD, $\widehat{\eta}$, can be directly calculated as 
\begin{eqnarray}\label{variance_wd}
	{\text{var}}(\widehat{\eta})={\text{var}}(\widehat{\tau}_{1}^{\text{UNADJ}})+{\text{var}}(\widehat{\tau}_{-1}^{\text{UNADJ}})-2{\text{cov}}(\widehat{\tau}_{1}^{\text{UNADJ}},\widehat{\tau}_{-1}^{\text{UNADJ}}).
\end{eqnarray}
Hence, variance estimates of WR and WD from the unadjusted estimators can be obtained from \eqref{variance_wr} and \eqref{variance_wd}, respectively. 
In Web Appendix A1-A2, we also present details about how to obtain consistent estimators for ${\text{var}}\left(\widehat{\tau}_{1}^{\text{UNADJ}}\right)$, ${\text{var}}\left(\widehat{\tau}_{-1}^{\text{UNADJ}}\right)$, and ${\text{cov}}\left(\widehat{\tau}_{1}^{\text{UNADJ}},\widehat{\tau}_{-1}^{\text{UNADJ}}\right)$.

\begin{remark}
\label{rmk:another_unadjusted_variance}
Our $\widehat{\tau}_{1}^{\text{UNADJ}}$ in \eqref{unadjust} is constructed following the general principles in \citet{Bebu2016} but differs from that existing estimator due to the weight $\omega_{ij}$. For both the variance expression in \citet{Bebu2016} and
\eqref{unadjusted_var}, we constrain the comparisons in the variance estimation to strictly different individuals (Web Appendix A1-A2), i.e., for both probabilities $\mathbb{P}\left[Y_i(1)>Y_j(0)~\&~Y_i(1)<Y_k(0)\right]$ and $\mathbb{P}\left[Y_i(1)>Y_j(0)~\&~Y_k(1)<Y_j(0)\right]$, we need $j\neq k$ and $i \neq k$. \cite{Ozenne2021} showed that this can lead to bias in small samples and potentially negative variance estimates for $\hat{\tau}_{1}^{\text{UNADJ}}$. They suggested a bias-reduced variance estimator based on first and second order H-decomposition of U-statistics that allows comparisons in the variance estimation to also come from the same individual (i.e., in the first probability, $j$ can be equal to $k$; and in the second probability, $i$ can be equal to $k$), which has less bias and can only be positive. Following their approach, we also provided a bias-reduced variance estimator for $\hat{\tau}_{1}^{\text{UNADJ}}$ in Web Appendix A3. By comparing the expressions given in Web Appendix A1 and A3, one can observe that the key difference lies in estimating the two probabilities $\mathbb{P}\left[Y_i(1)>Y_j(0)~\&~Y_i(1)<Y_k(0)\right]$ and $\mathbb{P}\left[Y_i(1)>Y_j(0)~\&~Y_k(1)<Y_j(0)\right]$ (by using \eqref{unadjusted_var} versus the method suggested by \cite{Ozenne2021}). We will also examine their performance by simulations in Section \ref{sec:simulation}.
\end{remark}

\section{Covariate-adjusted win statistics via propensity score weighting}\label{sec:weighting}
\subsection{The balancing weights}\label{subsec:balanced_weights}
In this Section, we introduce propensity score weighting methods for covariate-adjusted estimation of WRs and WDs. We first generalize the win estimand in \eqref{win_prob} to a class of weighted win estimands to motivate alternative weighting estimators. Specifically, we assume that the marginal density of $X$ (from the combined treatment and control groups) exists and is denoted by $f(X)$. Under the balancing weight framework \citep{Li2018,Li2019}, we can represent the target population density by $f(X)h(X)/C$, where $h(X)$ is a pre-specified tilting function, and $C$ is a normalization constant such that the density is well-defined and integrates to unity. The weighted win estimand for a given target population characterized by the tilting function is
\begin{eqnarray}\label{balanced_ate}
	\tau_{1}^h = \frac{\mathbb{E}\left\{h(X_i, X_j)\mathbb{I}\left[Y_i(1) > Y_j(0)\right]\right\}}{\mathbb{E}\left[h(X_i, X_j)\right]}.
\end{eqnarray}
Here $h(X_i,X_j)$ is a bivariate tilting function defined as the ratio of the covariate distribution density in the target population to that in the study sample. In principle, $h(X_i, X_j)$ is a symmetric function of the two individuals' covariates. In practice and throughout what follows, we will specify $h$ as a product of univariate tilting functions as $h(X_i, X_j) = h^*(X_i) h^*(X_j)$, where $h^*$ is the more familiar individual tilting function. This factorization satisfies the required symmetry and is natural given $h^*$ can be defined as a function of the propensity score, $e(x)=\mathbb{P}(Z=1|X)$.

In observational studies, propensity scores are usually unknown and must be estimated from observed data. Therefore, different $h(\cdot)$ corresponds to different populations and estimands. 
However, under randomization, for any baseline covariate value $X$, we have $e(X)=\mathbb{P}(Z=1)=\pi$.Thus, as long as $h(X_i,X_j)$ depends on covariates only through the propensity score, different $h(\cdot)$ correspond to the same target population and the weighted win estimand reduces to the unweighted win estimand: $\tau_{1}^h=\tau_{1}$. This unique feature provides the basis for considering alternative weighting estimators that perform implicit covariate adjustment without changing the target of inference \citep{zeng2021propensity}. 

\citet{zeng2021propensity} leveraged the class of balancing weights to perform covariate adjustment, with the weights defined as $\omega_1(X)=h(X)/e(X)$ and $\omega_0=h(X)/\left[1-e(X)\right]$ for the treatment and control arms, respectively. 
We generalize their approach to handle the bivariate tilting function $h(X_i,X_j)$. Specifically, we focus on the following H{\'a}jek-type estimator:
\begin{eqnarray}\label{balanced_est}
	\widehat{\tau}_{1}^h = \frac{\sum_{i=1}^n\sum_{j=1}^n\omega(X_i, X_j)Z_i(1-Z_j)\mathbb{I}(Y_i > Y_j)}{\sum_{i=1}^n\sum_{j=1}^nZ_i(1-Z_j)\omega(X_i, X_j)},
\end{eqnarray}
where $\omega(X_i, X_j) = \frac{\widehat{h}(X_i,X_j)}{\widehat{e}(X_i)[1-\widehat{e}(X_j)]}$. By the weak law of large numbers, we can show that $\widehat{\tau}_{1}^h$ is consistent to the target $\tau_{1}^h$ defined in \eqref{balanced_ate}:
\begin{eqnarray}\label{consistent}
	\widehat{\tau}_{1}^h\stackrel{p}{\longrightarrow}\tau_{1}^h,\qquad {\text{as}}~n \rightarrow \infty.
\end{eqnarray}
The results for $\tau_{-1}$ are analogous and are omitted for brevity.

\subsection{Covariate adjustment via inverse probability weighting and overlap weighting}\label{subsec:ipw_ow}
To leverage information from the baseline covariates, we first propose the following IPW estimator:
\begin{eqnarray}\label{ipw_est}
	\widehat{\tau}_{1}^{\text{IPW}} = \frac{\sum_{i=1}^n\sum_{j=1}^nZ_i(1-Z_j)\mathbb{I}(Y_i > Y_j)/\left\{\widehat{e}(X_i)[1-\widehat{e}(X_j)]\right\}}{\sum_{i=1}^n\sum_{j=1}^nZ_i(1-Z_j)/\left\{\widehat{e}(X_i)[1-\widehat{e}(X_j)]\right\}}.
\end{eqnarray}
Note that \eqref{ipw_est} is the special case of \eqref{balanced_est} by letting $\widehat{h}(X_i,X_j)=1$ and $\omega(X_i, X_j) = \frac{1}{\widehat{e}(X_i)[1-\widehat{e}(X_j)]}$. In practice, the propensity score $e(X)$ is usually estimated by a logistic regression model. Furthermore, it has been established that IPW can perform covariate adjustment and reduces the impact of chance imbalance, hence improving the precision over the unadjusted estimator, even when estimating a known propensity score rather than using the true propensity score $\pi$ \citep{shen2014inverse,williams2014variance}. \citet{Mao2018} explored an IPW method for a general class of causal estimands defined by a contrast function, which extends the familiar notion of average treatment effect. However, their IPW estimator is unstabilized Horvitz-Thompson estimator, and hence differs from the stabilized IPW estimator \eqref{ipw_est}. \cite{wang2025adjustedWR} developed three IPW
	estimators: IPW-average treatment effect (IPW-ATE), stabilized IPW-ATE
	(SIPW-ATE) and IPW-average treatment effect in the treated (IPW-ATT) for the win ratio estimand. Nevertheless, their stabilized IPW-ATE is not a H{\'a}jek-type estimator as \eqref{ipw_est} and they generally ignored the uncertainty for estimating the propensity score in variance estimation. More recently, \cite{buyse2025} briefly discussed covariate adjustment strategies including an IPW estimator for generalized pairwise comparison estimands, but they did not demonstrate how to conduct inference, nor did they provide a variance estimator. In contrast, we focus on a H{\'a}jek-type IPW estimator \eqref{ipw_est}, and will derive a consistent variance estimator accounting for the uncertainty in estimating the propensity score. 

Conventional IPW assigns a weight of $1/e(X)$ for those in the treatment arm and $1/\left[1-e(X)\right]$ for those in the control arm. Beyond IPW, OW assigns a weight of $1-e(X)$ for treated patients and a weight of $e(X)$ for control patients, and obviates the need for inverse weighting. In observational studies, it has been shown that OW retains the advantages of IPW while offering better finite-sample properties \citep{Mao2019,Li2019,Li2019address,Cheng2022address,cao2024using}. By choosing $\widehat{h}(X_i,X_j)=\widehat{e}(X_i)\widehat{e}(X_j)(1-\widehat{e}(X_i))(1-\widehat{e}(X_j))$ so that $\omega(X_i, X_j) = \widehat{e}(X_j)[1-\widehat{e}(X_i)]$ in \eqref{balanced_est}, we propose the H{\'a}jek-type estimator with OW as:
\begin{eqnarray}\label{ow_est}
	\widehat{\tau}_{1}^{\text{OW}} = \frac{\sum_{i=1}^n\sum_{j=1}^nZ_i(1-Z_j)[1-\widehat{e}(X_i)]\widehat{e}(X_j)\mathbb{I}(Y_i > Y_j)}{\sum_{i=1}^n\sum_{j=1}^nZ_i(1-Z_j)[1-\widehat{e}(X_i)]\widehat{e}(X_j)}.
\end{eqnarray}
In observational studies, OW has been shown to have the exact balance property and the minimum variance property \citep{Li2018}. That is, when $e(X)$ is estimated from a logistic regression, the propensity weighted covariate means are identical between the treatment and control groups (exact balance); under regularity conditions, OW leads to the most efficient estimator among the family of balancing weighting schemes for moment estimators (minimum asymptotic variance). The latter property also holds 
for modeling survival outcomes \citep{Cheng2022address,cao2024using}. For conventional average treatment effect estimands, \citet{zeng2021propensity} further proved that OW and IPW lead to the same asymptotic variance in RCTs. However, due to the exact balance property of OW \citep{Li2018}, \citet{zeng2021propensity} demonstrated that OW can effectively remove all chance imbalance when the data generating process follows a linear model with additive terms. Furthermore, their simulation studies have shown that OW is always at least as efficient as IPW in finite samples for covariate adjustment, and carries a particular efficiency advantage when the sample size is small. Due to these reasons, we also investigate the use of OW in the context of WR and WD estimands. 

\subsection{Asymptotic analysis of the covariate-adjusted win statistics}\label{subsec:asymptotic_weighting}
Building off the theory of \citet{Mao2018}, we now apply U-statistic theory to derive the asymptotic variance of IPW and OW estimators. This derivation serves two purposes: (1) to show that propensity score weighting generally reduces the asymptotic variance when estimating win estimand, and (2) to obtain consistent variance estimators for covariate-adjusted win estimators. For convenience, we denote the propensity score model $e(X)$ by $\mM_{e}$; here, we consider $\mM_{e}$ to be a logistic regression --- that is, $e(X)=\mathbb{P}(Z=1|X;\beta)=(1+\exp(-X^{\top} \beta))^{-1}$,	
where $\beta=(\beta_1,\dots,\beta_p)^{\top}$ is a finite $p$-dimensional parameter. Let $\widehat{\beta}$ be the maximum likelihood estimator for $\beta$ under $\mM_{e}$, then $\widehat{e}(X)=(1+\exp(-X^{\top} \widehat{\beta}))^{-1}$. Let $g: \mX\times \mX \rightarrow \mR$ be an arbitrary contrast function for two potential outcomes, where $\mX$ is the space in which the potential outcomes reside, and let $\nu_z$ denote the marginal distribution of $Y(z)$, for $z=0,1$. Then, the win estimand takes the general form of
\begin{eqnarray}\label{general_est}
	\tau^g=\int\int g(t,s)\nu_1(dt)\nu_0(ds),
\end{eqnarray}
based on a kernel $g$. When $g(t,s)=\mathbb{I}(t>s)$, $\tau^g=\tau_{1}$; similarly, when $g(t,s)=\mathbb{I}(t<s)$, $\tau^g=\tau_{-1}$. 
For generality, we focus on the asymptotic analysis for a general $g$ in what follows.   

Since both the IPW estimator $\widehat{\tau}_{1}^{\text{IPW}}$ defined in \eqref{ipw_est} and the OW estimator $\widehat{\tau}_{1}^{\text{OW}}$ defined in \eqref{ow_est} are special cases of the weighting estimator $\widehat{\tau}_{1}^{h}$ defined in \eqref{balanced_est}, we can rewrite the weighting estimator as
\begin{eqnarray*}
	\widehat{\tau}_{1}^{h}=
	{n \choose 2}^{-1}\sum_{i=1}^n\sum_{j>i}^n q^{h}(\mO_i,\mO_j;\widehat{\beta}),
\end{eqnarray*} 
where
\begin{equation*}
	q^{h}(\mO_i,\mO_j;\widehat{\beta})=\frac{n(n-1)\left\{Z_i(1-Z_j)\mathbb{I}(Y_i>Y_j)\omega(X_i,X_j)+Z_j(1-Z_i)\mathbb{I}(Y_i<Y_j)\omega(X_j,X_i)\right\}}{2\sum_{i=1}^n\sum_{j=1}^nZ_i(1-Z_j)\omega(X_i,X_j)}.
\end{equation*}
Then, for any $i \neq j$, we have $\mathbb{E}\left[q^{h}(\mO_i,\mO_j;\widehat{\beta})\right]=\tau_{1}^{h}$ under randomization. By the Uniform Law of Large Numbers for the U-process \citep{Arcones1993,VanderVaart1996} and consistency of $\widehat{\beta}$, we can establish the consistency of $\widehat{\tau}_{1}^{h}$. 

In Theorem \ref{prop1}, we show that: (1) the asymptotic variance of weighting estimator $\widehat{\tau}_{1}^{h}$ can be estimated by the second moment estimator for an estimated influence function; (2) both IPW and OW estimators have smaller asymptotic variance than the unadjusted estimator $\tau_{1}^{\text{UNADJ}}$. The proof of Theorem \ref{prop1} is provided in Web Appendix B.

\begin{theory}(Asymptotic results of propensity score weighting estimators)\label{prop1}\\
\emph{(a) The asymptotic variance of $n^{1/2}(\widehat{\tau}_{1}^{h}-\tau_{1}^{h})$ is ${\text {var}}\left(\phi^h(\mO)\right)$, where 
\begin{eqnarray}\label{inf_w}
\phi^h(\mO)=2\{q_1^{h}(\mO;\beta)-\tau_{1}^{h}\}+\frac{\partial}{\partial \beta^{\top}}(\mathbb{E} \times \mathbb{E} )q^{h}(\mO,\mO;\beta)\widetilde{l}_{\beta}(\mO;\beta),
\end{eqnarray}
is the influence function of $\widehat{\tau}_{1}^{h}$, $q_1^{h}(\mO;\beta)=\mathbb{E} q^{h}\left(\cdot,\mO;\beta\right)=\frac{1}{2}\left[\frac{Z}{e(X)}\int \mathbb{I}(Y>s)\nu_0(ds)+\frac{1-Z}{1-e(X)}\int \mathbb{I}(Y<t)\nu_1(dt)\right]$,
		and $\widetilde{l}_{\beta}\left(\mO;\beta\right)$ is the influence function for $\beta$ based on logistic regression. \\
(b) The influence function $\phi^h(\mO)$ in \eqref{inf_w} is equal to $\chi^h(\mO)$ minus its projection on the tangent space of $\mM_e$, where 
\begin{eqnarray*}
\chi^h(\mO)&=&2\{q_1^{h}(\mO;\beta)-\tau_{1}^{h}\}=\frac{Z}{e(X)}\int \mathbb{I}(Y>s)\nu_0(ds)+\frac{1-Z}{1-e(X)}\int \mathbb{I}(Y<t)\nu_1(dt)-2\tau_{1}^h.
\end{eqnarray*}
The detailed expressions of the tangent space of $\mM_e$ and the projection are provided in Web Appendix B.\\
(c) The unadjusted estimator $\widehat{\tau}_{1}^{\text{UNADJ}}$ also belongs to the same class of balancing weighting estimators $\widehat{\tau}_{1}^h$ defined in \eqref{balanced_est}, and the influence function of $\widehat{\tau}_{1}^{\text{UNADJ}}$ is equal to $\chi^h(\mO)$ with $e(X)=\pi$. So the asymptotic variance of unadjusted estimator $\tau_{1}^{\text{UNADJ}}$ is equal to ${\text{var}}\left(\chi^h(\mO)\right)$. }
\end{theory}

Theorem 1(a) is obtained from applying Lemma S1 of \citet{Mao2018} directly. By applying the same technique used in Remark S1 of \citet{Mao2018} and the semi-parametric efficiency theory \citep{tsiatis2006semiparametric}, we can derive the result in Theorem 1(b). Note that if we let $\widehat{h}(X_i,X_j)=\widehat{e}(X_i)[1-\widehat{e}(X_j)]$ and $\omega(X_i,X_j)=1$, then $\widehat{\tau}_{1}^h$ defined in \eqref{balanced_est} is equivalent to $\tau_{1}^{\text{UNADJ}}$ defined in \eqref{unadjust0}. From this viewpoint, the unadjusted estimator $\widehat{\tau}_{1}^{\text{UNADJ}}$ also belongs to the same class of balancing weighting estimators as $\widehat{\tau}_{1}^h$.
Therefore, we can construct the influence function of $\widehat{\tau}_{1}^{\text{UNADJ}}$ (denoted as $\phi^{\text{UNADJ}}(\mO)$ --- see Web Appendix B for details) in the format of \eqref{inf_w} but without the second term on the right hand side. Typically, for the unadjusted estimator $\widehat{\tau}_{1}^{\text{UNADJ}}$, $e(X)=\pi$ and $\tau_{1}^h=\tau_{1}$; then we can show $\phi^{\text{UNADJ}}(\mO)=\chi^h(\mO)$ and ${\text {var}}\left(\widehat{\tau}_{1}^{\text{UNADJ}}\right)={\text {var}}\left(\chi^h(\mO)\right)$. Theorem 1(c) shows that the influence function of a balancing weighting estimator can be expressed as the influence function of the unadjusted estimator minus its projection onto the nuisance tangent space, which in fact underlies the asymptotic variance reduction achieved through covariate adjustment by weighting in randomized trials. 

We note that Theorem 1 summarizes the large-sample properties of the IPW and OW estimators for win estimands in randomized trials, which explains the asymptotic variance reduction of weighting estimators theoretically, extending the findings 
demonstrated in \citet{shen2014inverse},  \citet{williams2014variance} and \citet{zeng2021propensity}. However, the weighting estimators derived in this article are based on U-statistics and our win estimand is more complex than the usual average treatment effect.

\begin{remark}
	The large-sample properties of the OW estimator derived in this article do not depend on equal randomization ($\pi=0.5$), which is different from that in \citet{zeng2021propensity}. The reason is that the OW estimator is constructed by a symmetric kernel function based on two weights, such that the expectation of the kernel function of OW is equal to that of IPW under large sample sizes (i.e., $\mathbb{E}(q^{\text{IPW}}(\mO_i,\mO_j;\widehat{\beta}))=\mathbb{E}(q^{\text{OW}}(\mO_i,\mO_j;\widehat{\beta}))$). Therefore, equal randomization is not necessary for the variance reduction property under either OW or IPW. 
\end{remark}

\subsection{Closed-form variance estimation}\label{subsec:closed_var_est}
By Theorem 1(a), ${\text{var}}(\widehat{\tau}_{1}^{h})$
can be consistently estimated as 
\begin{eqnarray}\label{var_west}
	n^{-2}\sum_{i=1}^n\left[2\left\{\widehat{q}_1^{h}(\mO_i; \widehat{\beta}_n)-\widehat{\tau}_{1}^{h}\right\}+\widetilde{l}^{\top}_{\beta}(Z_i,X_i;\widehat{\beta})A_{1}^{h} \right]^{\otimes 2},
\end{eqnarray}
where $\widehat{q}_1^{h}(\mO_i;\widehat{\beta})=(n-1)^{-1}\sum_{j \neq i}^nq^{h}\left(\mO_i,\mO_j;\widehat{\beta}\right)$ and
$A_{1}^{h}={n \choose 2}^{-1} \sum_{i=1}^n\sum_{j>i}^n \frac{\partial}{\partial \beta}q^{h}\left(\mO_i,\mO_j;\beta\right)|_{\beta=\widehat{\beta}}$,
$\widetilde{l}_{\beta}(Z,X;\widehat{\beta})$ is the influence function for $\widehat{\beta}$ and $a^{\otimes 2}=a a^{\top}$ for any vector $a$. Therefore, for $\widehat{\tau}_{1}^{\text{IPW}}$ (i.e., $\widehat{\tau}_{1}^{h}$ with $\omega(X_i,X_j)=1/\widehat{e}(X_i)[1-\widehat{e}(X_j)]$) and $\widehat{\tau}_{1}^{\text{OW}}$ (i.e., $\widehat{\tau}_{1}^{h}$ with $\omega(X_i, X_j) = \widehat{e}(X_j)[1-\widehat{e}(X_i)]$),we can obtain their consistent variance estimators based on \eqref{var_west}, and their explicit expressions are provided in Web Appendix C.

In fact, according to the result in Theorem 1(c) and \eqref{var_west}, the variance of $\widehat{\tau}_{1}^{\text{UNADJ}}$ can also be consistently estimated as the second moment estimator for an estimated influence function: 
\begin{eqnarray}\label{var_unadj}
	\widehat{\text {var}}\left(\widehat{\tau}_{1}^{\text{UNADJ}}\right) = n^{-2}\sum_{i=1}^n\left[2\left\{\widehat{q}_1^{\text{UNADJ}}(\mO_i)-\widehat{\tau}_{1}^{\text{UNADJ}}\right\} \right]^{\otimes 2},
\end{eqnarray}
where $\widehat{q}_1^{\text{UNADJ}}(\mO_i)=(n-1)^{-1}\sum_{j \neq i}^nq^{\text{UNADJ}}(\mO_i,\mO_j)$ and $q^{\text{UNADJ}}(\mO_i,\mO_j)=\frac{1}{2}\bigr[Z_i(1-Z_j)\mathbb{I}(Y_i>Y_j)+Z_j(1-Z_i)\mathbb{I}(Y_i<Y_j)\bigr]/\bigr[\frac{1}{n(n-1)}\sum_{i=1}^n\sum_{j=1}^nZ_i(1-Z_j)\bigr]$. In Section \ref{sec:simulation}, we will demonstrate that the variance estimators \eqref{var_unadj} and \eqref{unadjusted_var} perform similarly and are both valid when sample size is moderate, but that \eqref{var_unadj} generally has better small sample performance than \eqref{unadjusted_var}.

Finally, since we have established the asymptotic results for a general kernel function $g$, the results would be applicable for both the win and loss estimators and can inform the estimation of WR and WD through application of the delta method. Specifically, for the loss probability estimator 
$$\widehat{\tau}_{-1}^h = \frac{\sum_{i=1}^n\sum_{j=1}^n\omega(X_i, X_j)Z_i(1-Z_j)\mathbb{I}(Y_i < Y_j)}{\sum_{i=1}^n\sum_{j=1}^n\omega(X_i, X_j)Z_i(1-Z_j)},$$ 
we can obtain the corresponding variance estimators by simply changing the kernel function $g$.

Finally, denote $\phi_{1,i}^h=2\left\{\widehat{q}_{1}^{h}(\mO_i; \widehat{\beta})-\widehat{\tau}_{1}^{h}\right\}+\widetilde{l}^{\top}_{\beta}(Z_i,X_i;\widehat{\beta})A_{1}^{h}$ and $\phi_{-1,i}^h=2\left\{\widehat{q}_{-1}^{h}(\mO_i; \widehat{\beta})-\widehat{\tau}_{-1}^{h}\right\}+\widetilde{l}^{\top}_{\beta}(Z_i,X_i;\widehat{\beta})A_{-1}^{h}$ as the influence function of patient $i$ for the probability of win and probability of loss, respectively, where $\widehat{q}_{-1}^{h}(\mO_i; \widehat{\beta})$, $\widehat{\tau}_{-1}^{h}$ and $A_{-1}^{h}$   
are the corresponding estimators for the probability of loss --- see \eqref{var_west} for similar definitions. Then, the asymptotic covariance between $\widehat{\tau}_{1}^{h}$ and $\widehat{\tau}_{-1}^{h}$ can be computed as ${\text {cov}}\left(\widehat{\tau}_{1}^{h},\widehat{\tau}_{-1}^{h}\right)=n^{-1}\sum_{i=1}^n \phi_{1,i}^h\phi_{-1,i}^h$. By applying formulas \eqref{variance_wr} and \eqref{variance_wd}, we can then assemble the asymptotic variance and covariance expressions to obtain final variance estimators for WR ($\theta$) and WD ($\eta$) based on IPW and OW.  

\section{Covariate-adjusted win statistics via augmented weighting}\label{sec:augmented}
\subsection{Formulating an augmented IPW estimator with outcome regression}
Both IPW and OW estimators only use a working propensity score model to adjust for chance imbalance. Hence, intuitively, potentially more efficient estimators could be constructed by leveraging the associations between baseline covariates and the potential outcomes through an additional outcome regression model, that is, the augmented weighting methods. For observational studies, \cite{Mao2018} proposed locally efficient and doubly robust estimators for general causal estimands based on contrast functions, formulated using Horvitz-Thompson weighting. \cite{Mao2019} extended this framework to study augmented weighting estimators, including augmented IPW and OW variants for weighted average treatment effects, and derived closed-form variance expressions. More recently, \cite{buyse2025} discussed an augmented IPW estimator for generalized pairwise comparison estimands, using Horvitz--Thompson weights. In what follows, we first review the formulation of the augmented IPW estimator discussed in \citet{Mao2018}. Building on this estimator, we then extend the approach to construct stabilized augmented weighting estimators for the win estimand allowing for the general class of balancing weights.

Let $\nu_z(\cdot|X;\alpha)$ denote the conditional laws (i.e., conditional density for continuous random variable or conditional probability for discrete random variable) of $Y(Z)$ given $X$, indexed by a Euclidean parameter $\alpha$, and let $\mathcal{M}_{\mu}$ denote the intersection of these two marginal models. For notational simplicity, we assume both $\mathcal{M}_{\mu}$ and $\mathcal{M}_{e}$ use the same set of covariates, though in practice these may differ. Given two independent individuals with $X_i$ and $X_j$, an outcome regression component can be defined as
\begin{eqnarray*}
	\mu(X_i,X_j;\alpha) = \mathbb{E}[g\{Y_i(1),Y_j(0)\}| X_i, X_j; \alpha]= \iint g(t,s)\nu_1(dt|X_i; \alpha)\nu_0(ds|X_j; \alpha).
\end{eqnarray*}
After finding the maximum likelihood estimator $\widehat{\alpha}$ for $\alpha$ under $\mM_{\mu}$ based on $(Y_i,Z_i,X_i)(i=1,\dots,n)$, and obtaining $\widehat{\mu}(X_i,X_j)=\mu(X_i,X_j;\widehat{\alpha})$, the augmented IPW estimator for $\tau^g$ in \citet{Mao2018} is given by
\begin{eqnarray*}
	\widehat{\tau}^{\text{MAO}}_{1} = \binom{n}{2}^{-1} \sum_{i=1}^n\sum_{j>i}^n r^{\text{MAO}} (\mathcal{O}_i, \mathcal{O}_j; \widehat{\beta}, \widehat{\alpha}),
\end{eqnarray*}
where
\begin{eqnarray*}
	r^{\text{MAO}}(\mathcal{O}_i, \mathcal{O}_j; \widehat{\beta}, \widehat{\alpha}) &= &
	2^{-1} \left(\frac{Z_i(1-Z_j)g(Y_i,Y_j)}{\widehat{e}(X_i)\{1-\widehat{e}(X_j)\}} - \left[\frac{Z_i(1-Z_j)}{\widehat{e}(X_i)\{1-\widehat{e}(X_j)\}}-1\right]\widehat{\mu}(X_i,X_j) \right. \\
	&& \left. 
	+ \frac{Z_j(1-Z_i)g(Y_j,Y_i)}{\widehat{e}(X_j)\{1-\widehat{e}(X_i)\}} - \left[\frac{Z_j(1-Z_i)}{\widehat{e}(X_j)\{1-\widehat{e}(X_i)\}}-1\right]\widehat{\mu}(X_j,X_i) \right).
\end{eqnarray*}
Alternatively, we can rewrite the above augmented weighting estimator as follows:
\begin{eqnarray*}
	\widehat{\tau}_{1}^{\text{MAO}} 
	&= & \frac{1}{n(n-1)} \sum_{i=1}^n \sum_{j>i}^n  \left\{ \frac{Z_i(1-Z_j)[g(Y_i, Y_j)-\widehat{\mu}(X_i,X_j)]}{\widehat{e}(X_i)(1-\widehat{e}(X_j))}+\frac{Z_j(1-Z_i)[g(Y_j, Y_i)-\widehat{\mu}(X_j,X_i)]}{\widehat{e}(X_j)(1-\widehat{e}(X_i))}\right\}\\
	&&+ \frac{1}{n(n-1)} \sum_{i=1}^n \sum_{j>i}^n [\widehat{\mu}(X_i, X_j) + \widehat{\mu}(X_j, X_i)].
\end{eqnarray*}

Specifically, for an ordinal outcome $Y$, let $p_{l}^z(X;\alpha)$ denote the model-based conditional probability that $Y(Z) = l$ $(l=1,2,\dots,L)$ given baseline variables $X$ and parameter $\alpha$, then $p_{l}^z(X;\alpha)=\nu_z(Y(Z)=l|X;\alpha)$. Assume that individual $i$ belongs to the treatment group ($Z=1$) and individual $j$ to the control group ($Z=0$). Then, \citet{Mao2018} has shown that
\begin{eqnarray*}
	\mu(X_i,X_j;\alpha) &=& \sum_{l=1}^L \sum_{l'=1}^{l-1} p_{l}^1(X_i;\alpha)p_{l'}^0(X_j;\alpha),\qquad \mu(X_j,X_i;\alpha) = \sum_{l=1}^L \sum_{l'=l+1}^{L} p_{l}^1(X_i;\alpha)p_{l'}^0(X_j;\alpha).
\end{eqnarray*}
In our simulation study, we primarily focus on ordinal logistic regression to model $\mu(\cdot,\cdot;\alpha)$ and to obtain fitted estimates using the \texttt{polr} function from the R package \texttt{MASS} \citep{MASS_2002}. In the \texttt{polr} function, the ordinal logistic regression model for an ordinal outcome with $L$ categories is parameterized as
\begin{eqnarray*}
	{\text{logit}}(P(Y\leq l|X))=\alpha_{l0}-\xi^{\top} X \Leftrightarrow P(Y\leq l|X)=\frac{\exp(\alpha_{l0}-\xi^{\top} X)}{1+\exp(\alpha_{l0}-\xi^{\top} X)}.
\end{eqnarray*}
where $l=1,2,\dots,L-1$, $\xi=(\xi_1,\dots,\xi_p)^{\top}$ and $\alpha=(\alpha_{10},\dots,\alpha_{L-1,0},\xi^{\top})^{\top}$. 
Then, we have
\[
p_{l}(X;\alpha)=P(Y=l) =
\begin{cases}
	P(Y \leq 1|X)=P(Y = 1|X),& \text{if } Y=l=1\\
	P(Y \leq l|X)-P(Y \leq l-1|X)& \text{if } 1<Y=l<L\\
	1-P(Y \leq L-1|X),              & \text{if } Y=l=L.
\end{cases}
\]
However, the augmented IPW estimator is fairly general and in theory can be used with any working outcome models, including more flexible ordinal regression models that permit the odds ratios for covariates to possibly change with cut-point. 

\subsection{Covariate adjustment via augmented inverse probability weighting and overlap weighting}
It is worth noting that when using the augmented IPW estimator $\widehat{\tau}_{1}^{\text{MAO}}$, the pairwise comparison concerning two individuals from the same treatment group (i.e., $Z_i = Z_j = 1$ or $Z_i = Z_j = 0$) simplifies to $r^{\text{MAO}}(\mathcal{O}_i, \mathcal{O}_j; \widehat{\beta}, \widehat{\alpha}) = \bigr[\widehat{\mu}(X_i,X_j)+\widehat{\mu}(X_j,X_i)\bigr]/2$. Therefore, the estimator $\widehat{\tau}^{\text{MAO}}_{1}$ can be reexpressed as
\begin{eqnarray*}
	\tilde{\tau}_{1}^{\text{MAO}} & =& \frac{1}{n(n-1)} \sum_{i=1}^n \sum_{j\neq i}^n \frac{Z_i(1-Z_j)[g(Y_i, Y_j)-\widehat{\mu}(X_i,X_j)]}{\widehat{e}(X_i)(1-\widehat{e}(X_j))} + \frac{1}{n(n-1)}\sum_{i=1}^n \sum_{j\neq i}^n  \widehat{\mu}(X_i, X_j).
\end{eqnarray*}

For an ordinal outcome, the kernel function used to define the win probability $\tau_{1}$ in \eqref{win_prob} is given by $g(Y_i,Y_j)=\mathbb{I}(Y_i>Y_j)$. Since the estimator $\tilde{\tau}_{1}^{\text{MAO}}$ is not a H{\'a}jek-type estimator, we incorporate a tilting function $\widehat{h}(X_i, X_j)$ and stabilized weights to propose the following augmented weighting (AW) estimator:
\begin{eqnarray}\label{ah_est}
	\widehat{\tau}_{1}^{\text{AW}} & =& \frac{\sum_{i=1}^n \sum_{j\neq i}^n  Z_i(1-Z_j)\omega(X_i, X_j)\{\mathbb{I}(Y_i>Y_j)-\widehat{\mu}(X_i,X_j)\}}
	{\sum_{i=1}^n \sum_{j\neq i}^n  Z_i(1-Z_j)\omega(X_i, X_j)}+ \frac{\sum_{i=1}^n \sum_{j\neq i}^n  \widehat{h}(X_i, X_j)\widehat{\mu}(X_i,X_j)}
	{\sum_{i=1}^n \sum_{j\neq i}^n  \widehat{h}(X_i, X_j)},
\end{eqnarray}
where
$\omega(X_i, X_j) = \widehat{h}(X_i, X_j)/\{\widehat{e}(X_i)(1-\widehat{e}(X_j))\}$. In the case where IPW is used, the tilting function is $\widehat{h}(X_i, X_j) = 1$ and the weight function is defined as
$\omega(X_i, X_j) = \{\widehat{e}(X_i)(1-\widehat{e}(X_j))\}^{-1}$. This leads to the following augmented inverse probability weighting (AIPW) estimator:
\begin{eqnarray*}\label{AIPW}
	\widehat{\tau}_{1}^{\text{AIPW}} &=&\frac{\sum_{i=1}^n \sum_{j\neq i}^n Z_i(1-Z_j)\{\mathbb{I}(Y_i>Y_j)-\widehat{\mu}(X_i,X_j)\}/[\widehat{e}(X_i)(1-\widehat{e}(X_j))]}{\sum_{i=1}^n \sum_{j\neq i}^n Z_i(1-Z_j)/[\widehat{e}(X_i)(1-\widehat{e}(X_j))]}+\frac{\sum_{i=1}^n \sum_{j\neq i}^n \widehat{\mu}(X_i,X_j)}{n(n-1)}.
\end{eqnarray*}
Alternatively, by setting the tilting function to $\widehat{h}(X_i,X_j) = \widehat{e}(X_i)\widehat{e}(X_j)(1-\widehat{e}(X_i))(1-\widehat{e}(X_j))$, we have $\omega(X_i, X_j) = \{1-\widehat{e}(X_i)\}\widehat{e}(X_j)$ and the following augmented overlap weighting (AOW) estimator is defined as:
\begin{eqnarray*}\label{AOW}
	\widehat{\tau}_{1}^{\text{AOW}} 
	&=&\frac{\sum_{i=1}^n \sum_{j\neq i}^n Z_i(1-Z_j)\widehat{e}(X_j)(1-\widehat{e}(X_i))\{\mathbb{I}(Y_i>Y_j)-\widehat{\mu}(X_i,X_j)\}}{\sum_{i=1}^n \sum_{j\neq i}^n Z_i(1-Z_j)[1-\widehat{e}(X_i)]\widehat{e}(X_j)}\\
	&&\\
	&&+\frac{\sum_{i=1}^n \sum_{j\neq i}^n \widehat{e}(X_i)\widehat{e}(X_j)(1-\widehat{e}(X_i))(1-\widehat{e}(X_j))\widehat{\mu}(X_i,X_j)}{\sum_{i=1}^n \sum_{j\neq i}^n \widehat{e}(X_i)\widehat{e}(X_j)[1-\widehat{e}(X_i)][1-\widehat{e}(X_j)]}. \nonumber
\end{eqnarray*}

\subsection{Asymptotic analysis of the AIPW and AOW estimators for win estimands}\label{subsec:asymptotic_augweighting}
To derive the variance of $	\widehat{\tau}_{1}^{\text{AW}}$ as defined in \eqref{ah_est}, we first express it in the form of a U-statistic: 
\begin{eqnarray*}
	\widehat{\tau}_{1}^{\text{AW}} & =& \frac{\{n(n-1)\}^{-1}\sum_{i=1}^n \sum_{j\neq i}^n  Z_i(1-Z_j)\omega(X_i, X_j)\{\mathbb{I}(Y_i>Y_j)-\widehat{\mu}(X_i,X_j)\}}
	{\{n(n-1)\}^{-1}\sum_{i=1}^n \sum_{j\neq i}^n Z_i(1-Z_j)\omega(X_i, X_j)}\\
	&&+ \frac{\{n(n-1)\}^{-1}\sum_{i=1}^n \sum_{j\neq i}^n  \widehat{h}(X_i, X_j)\widehat{\mu}(X_i,X_j)}
	{\{n(n-1)\}^{-1}\sum_{i=1}^n \sum_{j\neq i}^n  \widehat{h}(X_i, X_j)}={n \choose 2}^{-1}\sum_{i=1}^n\sum_{j>i}^nr^{h}(\mO_i,\mO_j;\widehat{\beta},\widehat{\alpha}),
\end{eqnarray*}
where \begin{eqnarray*}
	r^{h}(\mO_i,\mO_j;\widehat{\beta},\widehat{\alpha})
	&=&\frac{2^{-1}\{Z_i(1-Z_j)\omega(X_i,X_j)[\mathbb{I}(Y_i>Y_j)-\widehat{\mu}(X_i,X_j)]\}}{\{n(n-1)\}^{-1}\sum_{i=1}^n \sum_{j\neq i}^n Z_i(1-Z_j)\omega(X_i,X_j)}\\
	&&+\frac{2^{-1}\{Z_j(1-Z_i)\omega(X_j,X_i)[\mathbb{I}(Y_i<Y_j)-\widehat{\mu}(X_j,X_i)]\}}{\{n(n-1)\}^{-1}\sum_{i=1}^n \sum_{j\neq i}^n Z_i(1-Z_j)\omega(X_i,X_j)}+\frac{2^{-1}\widehat{h}(X_i,X_j)[\widehat{\mu}(X_i,X_j)+\widehat{\mu}(X_j,X_i)]}{\{n(n-1)\}^{-1}\sum_{i=1}^n \sum_{j\neq i}^n \widehat{h}(X_i,X_j)}.
\end{eqnarray*} 

By the uniform law of large numbers for U-processes \citep{Arcones1993,VanderVaart1996} and the consistency of both $\widehat{\beta}$ and $\widehat{\alpha}$, $\widehat{\tau}_{1}^{\text{AW}}$ is consistent when either $\mM_{e}$ or $\mM_{\mu}$ is correctly specified. Theorem 2 below provides the asymptotic variance of the augmented weighting estimator and establishes the local efficiency and robustness of $\widehat{\tau}_{1}^{\text{AW}}$. 

\begin{theory}(Asymptotic results of augmented weighting estimators)\label{prop2}\\ 
	\emph{(a)  The asymptotic variance of $n^{1/2}(\widehat{\tau}_{1}^{\text{AW}}-\tau_{1}^{h})$ is ${\text {var}}\left(\psi^h(\mO)\right)$, where 
		\begin{eqnarray*}\label{inf_augw}
			\psi^h(\mO)=2\{r_{1}^{h}(\mO;\beta,\alpha)-\tau_{1}^{h}\}+\frac{\partial}{\partial \beta^{\top}}(\mathbb{E} \times \mathbb{E})r^{h}(\mO,\mO;\beta,\alpha)\widetilde{l}_{\beta}(\mO;\beta)+\frac{\partial}{\partial \alpha^{\top}}(\mathbb{E} \times \mathbb{E})r^{h}(\mO,\mO;\beta,\alpha)\widetilde{l}_{\alpha}(\mO;\alpha),
		\end{eqnarray*}
		is the influence function of $\widehat{\tau}_{1}^{\text{AW}}$ with
		\begin{eqnarray*}
			r_{1}^{h}(\mO;\beta,\alpha)=\mathbb{E} r^{h}\left(\cdot,\mO;\beta,\alpha\right)&=&\frac{1}{2}\frac{Z}{e(X)}\left\{\int \mathbb{I}(Y>s)\nu_0(ds)-E\left(\int \mathbb{I}(Y(1)>s)\nu_0(ds)|X\right)\right\}\\
			&&+\frac{1}{2}\frac{1-Z}{1-e(X)}\left\{\int \mathbb{I}(Y<t)\nu_1(dt)-E\left(\int \mathbb{I}(Y(1)<t)\nu_1(dt)|X\right)\right\}\\
			&&+\frac{1}{2}\left[E\left(\int \mathbb{I}(Y(1)>s)\nu_0(ds)|X\right)+E\left(\int \mathbb{I}(Y(0)<t)\nu_1(dt)|X\right)\right],
		\end{eqnarray*}
		and $\widetilde{l}_{\alpha}\left(\mO;\alpha\right)$ is the influence function for $\alpha$. \\
		(b) Under randomization, both the AIPW and AOW estimators, $\widehat{\tau}_{1}^{\text{AIPW}}$ and $\widehat{\tau}_{1}^{\text{AOW}}$ are robust in the sense that they are always consistent to $\tau_1^h$ regardless of whether the ordinal outcome regression model $\mM_{\mu}$ is correctly specified. When the ordinal outcome regression model $\mM_{\mu}$ is correctly specified, both estimators are locally efficient.}
\end{theory}

The consistent variance estimators for $\widehat{\tau}_{1}^{\text{AW}}$ can be constructed similarly to those for $\widehat{\tau}_{1}^{h}$ in \eqref{var_west}. According to lemma S1 of \citet{Mao2018}, we have $2[E\{r^{h}(\mO_i,\mO_j;\beta,\alpha)|\mO_i\}-\tau_1^h]=\psi^h(\mO_i)$, then by Hoeffding's decomposition theorem \citep{Hoeffding1948}, the remainder of the proof for Theorem 2(b) follows the same steps as for the average treatment effect discussed in Chapter 13 of \citet{tsiatis2006semiparametric}. Typically, when $\mM_{\mu}$ is correct, $\widehat{\alpha}$ has no effect on the asymptotic distribution of $\widehat{\tau}_{1}^{\text{AW}}$. In other words, $\widetilde{l}_{\alpha}\left(\mO;\alpha\right)$ has no contribution to $\psi^h(\mO)$ and $\psi^h(\mO)$ can be rewritten as
\begin{eqnarray*}
	\psi^h(\mO)=2\{r_{1}^{h}(\mO;\beta,\alpha)-\tau_{1}^{h}\}+\frac{\partial}{\partial \beta^{\top}}(\mathbb{E} \times \mathbb{E})r^{h}(\mO,\mO;\beta,\alpha)\widetilde{l}_{\beta}(\mO;\beta).
\end{eqnarray*}
Note that for $\tau_{1}^{\text{UNADJ}}$, $\pi=e(X)$, so $2\{r_{1}^{h}(\mO;\beta,\alpha)-\tau_{1}^{h}\}=\chi^h(\mO)$ (i.e., the influence function of $\tau_{1}^{\text{UNADJ}}$). By using the same technique in Theorem 1(b), we can also show that $\psi^h(\mO)$ is equal to $\chi^h(\mO)$ minus its projection on the tangent space of $\mM_e$. Therefore, the asymptotic variance reduction by augmented weighting estimators is guaranteed when the outcome model is correctly specified, a finding that is consistent to that in \citet{Vermeulen2015increasing} for studying covariate-adjusted Mann-Whitney U tests.

Finally, based on Theorem \ref{prop2}, we can also obtain the closed-form variance estimators of $\widehat{\tau}_{1}^{\text{AIPW}}$ and $\widehat{\tau}_{1}^{\text{AOW}}$; the technical details and full expressions are provided in Web Appendix D. Similarly, we can apply the procedure described in the penultimate paragraph of Section \ref{subsec:closed_var_est} to derive the covariance between the win probability $\widehat{\tau}_{1}^{\text{AW}}$ and loss probability $\widehat{\tau}_{-1}^{\text{AW}}$. This allows us to estimate the variance of $\widehat{\theta}$ using formula \eqref{variance_wr}, and the variance of $\widehat{\eta}$ using formula \eqref{variance_wd}, for both the AIPW and AOW estimators. To assist practical applications, we implement the unadjusted and all covariate-adjusted point and variance estimators in an R package \protect\texttt{winPSW}.

\section{Simulation studies}\label{sec:simulation}
To investigate the finite-sample performance of unadjusted, weighting and augmented weighting estimators, we conduct a series of simulation studies under several data generation processes. 
Our overarching goal is to empirically (1) examine the accuracy of the point estimators for WR and WD with ordinal outcomes; (2) validate the variance estimators of the unadjusted, weighting and augmented weighting methods developed in \eqref{unadjusted_var} or \eqref{var_unadj}, (A.8), (A.9) in Web Appendix C, (A.13) and (A.14) in Web Appendix D, respectively; (3) explore whether the weighting and augmented weighting estimators can lead to finite-sample efficiency gains compared with the unadjusted estimator. Through these simulation studies, we also seek to confirm the model robustness property of augmented weighting estimators under a mis-specified working ordinal outcome model.  

\subsection{Simulation design}\label{subsec:simdesign}
We generate $p=6$ baseline covariates (three continuous and three binary covariates): $X_1\sim N(1,0.3^2)$, $X_2 \sim N(0.9,0.4^2)$, $X_3 \sim N(0.8,0.5^2)$, $X_4 \sim \text{Bernoulli}(0.75)$, $X_5 \sim \text{Bernoulli}(0.5)$, and $X_6 \sim \text{Bernoulli}(0.25)$. The treatment indicator is generated from $Z \sim \text{Bernoulli}(\pi)$ by fixing the randomization probability $\pi$. For illustrative purposes, we assume the ordinal outcome has three levels and that a higher level indicates a more favorable outcome. To examine the model robustness feature of our covariate-adjusted estimators, we consider two types of potential outcomes data generating processes: a quadratic association cumulative logit model, and a covariate interaction cumulative logit model. Specifically, based on baseline covariates $X_i=(X_{i1},X_{i2},\dots,X_{i6})^{\top}$, we generate the potential outcomes from the following quadratic model: for $z=0,1$,
\begin{eqnarray}\label{quadratic}
	Y_i(z) \sim \left\{
	\begin{aligned}
		{\text{logit}}[\mathbb{P}(Y_i(z)\geq 2)]=\alpha_{10}+\sum_{j=1}^6\gamma_jX_{ij}+\sum_{j=1}^6\gamma_{quad,j}X_{i,j}^2+\delta z,\\
		{\text{logit}}[\mathbb{P}(Y_i(z)\geq 3)]=\alpha_{20}+\sum_{j=1}^6\gamma_jX_{ij}+\sum_{j=1}^6\gamma_{quad,j}X_{i,j}^2+\delta z, 
	\end{aligned} \right\}
\end{eqnarray}
where $\alpha_{10}=1$, $\alpha_{20}=0.05$, $\delta=1$, $\gamma=(\gamma_1,\gamma_2,\gamma_3,\gamma_4,\gamma_5,\gamma_6)^{\top}=(1,-1,1,-1,1,-1)^{\top}\times (0.5+0.5z)$ and $\gamma_{quad}=(\gamma_{quad,1},\gamma_{quad,2},\gamma_{quad,3},\gamma_{quad,4},\gamma_{quad,5},\gamma_{quad,6})^{\top}=2\gamma$. Therefore, the parameter $\alpha$ in outcome $\mu(X_i,X_j;\alpha)$ can be characterized as $\alpha=(\alpha_{10},\alpha_{20},\gamma^{\top},\gamma^{\top}_{quad},\delta)^{\top}$. In addition, we consider the following interaction model to generate the potential outcomes: for $z=0,1$,
\begin{eqnarray}\label{interaction}
	Y_i(z) \sim \left\{
	\begin{aligned}
		{\text{logit}}[\mathbb{P}(Y_i(z)\geq 2)]=\alpha_{10}+\sum_{j=1}^6\gamma_jX_{ij}+\sum_{j=1}^6\sum_{j<k}^6\gamma_{int,jk}X_{i,j} X_{i,k}+\delta z,\\
		{\text{logit}}[\mathbb{P}(Y_i(z)\geq 3)]=\alpha_{20}+\sum_{j=1}^6\gamma_jX_{ij}+\sum_{j=1}^6\sum_{j<k}^6\gamma_{int,jk}X_{i,j} X_{i,k}+\delta z, 
	\end{aligned} \right\}
\end{eqnarray}
where $\gamma_{int,jk}=\mathbb{I}(\gamma_j\times \gamma_k>0)\times (\gamma_j+\gamma_k)+\mathbb{I}(\gamma_j\times \gamma_k<0)\times 0.25\times (\gamma_j\times \gamma_k)$ and $\delta=2$. Values of other parameters $\alpha_{10}$, $\alpha_{20}$ and $\gamma$ are the same as those in quadratic model \eqref{quadratic}. The observed outcome is set to be $Y_i=Y_i(Z_i)=Z_iY_i(1)+(1-Z_i)Y_i(0)$. For the randomization probability $\pi$, two values are considered: $\pi=0.5$ and $\pi=0.7$, representing a balanced assignment and an unbalanced assignment.

Under each simulation scenario, true values of WR $\theta$ and WD $\eta$ are determined using Monte Carlo approximation based on the simulated potential outcomes. We compare the unadjusted estimator $\widehat{\tau}^{\text{UNADJ}}$ (UNADJ), the IPW estimator $\widehat{\tau}^{\text{IPW}}$ (IPW), the OW estimator $\widehat{\tau}^{\text{OW}}$ (OW), the augmented IPW estimator $\widehat{\tau}^{\text{AIPW}}$ (AIPW) and the augmented OW estimator $\widehat{\tau}^{\text{AOW}}$ (AOW) for WR and WD. For weighting estimators and augmented weighting estimators, the propensity score is estimated by logistic regression including six baseline covariates as linear terms. For the augmented weighting estimators, we keep the same propensity score model specification, and the working outcome regression model is specified as an ordinal logistic regression including the six baseline covariates as linear terms plus quadratic terms or plus interaction terms.  

When evaluating the accuracy of the point estimator, we generate the observed data $\mO_i=(Y_i,Z_i,X_i)$ with three different sample sizes $n \in \{200,300,400\}$ under both balanced and unbalanced designs based on the quadratic association model \eqref{quadratic} and the covariate interaction model \eqref{interaction}, and calculate the relative bias (bias divided by the true estimand value times 100). To explore the model robustness property of the augmented weighting estimators, we also explore the results when we omit quadratic terms $X^2_{i,j}$ and interaction terms $X_{i,j} X_{i,k}$ ($j\neq k$) when specifying the ordinal outcome logistic regression. We refer to this scenario as ``model mis-specification'' and corresponding augmented IPW and augmented OW as AIPW-mis and AOW-mis, respectively.  

Beyond the accuracy of the point estimators, we also investigate the accuracy of the proposed variance estimators and the empirical coverage of the 95\% confidence intervals (based on normal approximation). Specifically, for $k=1$ or $k=-1$, the confidence intervals of $\widehat{\tau}_{k}^{\text{UNADJ}}$, $\widehat{\tau}_{k}^{\text{IPW}}$, $\widehat{\tau}_{k}^{\text{OW}}$, $\widehat{\tau}_{k}^{\text{AIPW}}$, $\widehat{\tau}_{k}^{\text{AOW}}$ are constructed based on \eqref{var_unadj}, (A.8), (A.9), (A.13) and (A.14) in Web Appendix, respectively. Then, based on \eqref{variance_wr} and \eqref{variance_wd}, we can obtain confidence intervals of $\widehat{\theta}$ and $\widehat{\eta}$. To quantify the accuracy of the estimated variance of each estimator, we report the ratio between the average variance estimates and the Monte Carlo variance (considered as the truth) of each estimator, with a ratio close to 1 indicating adequate performance. In addition, we also compare the variance estimator of $\widehat{\tau}_{1}^{\text{UNADJ}}$ based on \eqref{unadjusted_var} (the original \cite{Bebu2016} approach), the bias-reduced version provided in Web Appendix A3, and \eqref{var_unadj} (our alternative variance estimation approach). We report relative bias (\%), Monte Carlo standard deviation
(MCSD), average estimated standard error (AESE) based on \eqref{unadjusted_var} and \eqref{var_unadj}, and corresponding coverage probability (CP)
of the 95\% confidence intervals for the WR and WD. 

To illustrate the finite-sample efficiency gains under different sample sizes, we designed a separate set of simulations based on both the quadratic model \eqref{quadratic} and the interaction model \eqref{interaction}, where the total sample sizes $n$ vary from 60 to 600 to mimic small and large sample size scenarios. We report the efficiency results for estimating WR and WD based on the Monte Carlo variance across all simulation scenarios. The relative efficiency of an estimator $\widehat{\theta}$ (or $\widehat{\eta}$) is defined as ${{\text{var}}(\widehat{\theta}^{\text{UNADJ}})}/{{\text{var}}(\widehat{\theta})}$ (or ${{\text{var}}(\widehat{\eta}^{\text{UNADJ}})}/{{\text{var}}(\widehat{\eta})}$), i.e., the ratio between the Monte Carlo variance of the unadjusted estimator and that of the estimator under consideration. A higher relative efficiency indicates that the estimator is more efficient than the unadjusted estimator, and hence covariate adjustment is beneficial. We also report the relative efficiency comparisons between the augmented weighting estimators and the unadjusted estimator when the working outcome regression model is mis-specified (i.e., the quadratic terms and interaction terms are omitted when fitting the outcome $Y$ by an ordinal logistic regression). For each scenario, we simulate 1000 data replicates and the results are summarized based on these replicates.

\subsection{Simulation results}\label{subsec:simresults}
\subsubsection{Point estimation}
\begin{center}
[Table \ref{tb:RB} about here.]
\end{center}

The relative bias of UNADJ, IPW, OW, AIPW, AOW, AIPW-mis and AOW-mis under each scenario are summarized in Table \ref{tb:RB}. For the WR, we notice slight relative bias from all estimators even when the sample size is as small as 200. When the sample size increases to 400, all relative biases for estimating WR from weighting and augmented weighting estimators are negligible (often less than 1\%).
Furthermore, under each scenario, the bias of the WR from the unadjusted estimator is usually the largest, suggesting that covariate adjustment may reduce finite-sample bias. For estimating the WD, the relative biases from all estimators remain negligible and the bias from the unadjusted estimator is again slightly larger than that from the covariate-adjusted estimators. However, under an unbalanced design and when sample size is as small as 200, the bias from the unadjusted estimator can be slightly smaller than that from the covariate-adjusted estimators. Comparing biases of AIPW-mis and AOW-mis with those of AIPW and AOW, we find that the relative bias of AIPW and AOW is not substantially different regardless of whether the outcome model is correctly specified. This highlights the outcome model robustness property of the augmented weighting estimators.

\subsubsection{Variance and interval estimation}
\begin{center}
	[Table \ref{tab:variance_and_coverage} about here.]
\end{center}

Table \ref{tab:variance_and_coverage} reports the results on the accuracy of the proposed variance estimators and the empirical coverage rates of the associated 95\% confidence interval estimators for WR and WD under both balanced and unbalanced designs. Overall, we find that the average estimated variance to Monte Carlo variance ratio is generally close to 1 under all scenarios, especially when the sample size reaches $n=400$. This pattern clearly demonstrates that all of the proposed variance estimators are generally accurate in capturing the true variability of the covariate-adjusted estimators. Noticeably, even when sample size reaches 400, the ratio between the average estimated variance to Monte Carlo variance from both AIPW and AOW under the covariate interaction data generating model is generally less than 0.9; this is especially the case for AIPW under the unbalanced design, where this ratio drops below 0.8. Upon closer inspection, we find that when fitting both AIPW and AOW under the correct outcome model, there are a handful of baseline covariates (6 baseline covariates plus 15 interaction terms) in the fitted outcome model. Hence, the increasing number of coefficients that need to be estimated compromises the finite-sample convergence of the variance estimator to the truth. When the misspecified outcome model only includes 6 baseline covariates (AIPW-mis and AOW-mis), the ratio between the average estimated variance to Monte Carlo variance are closer to 1. This result shows that for augmented weighting estimators, if the sample size is not large, including too many covariates in the outcome model may lead to underestimation of variance. Finally, as the sample size increases, coverage rates generally improve across all estimators, reflecting better estimation of the true variance. Particularly, our proposed variance estimators accurately capture the true variability of AIPW-mis and AOW-mis even under unbalanced designs.

We also explore the performance of the proposed variance estimator against the original \cite{Bebu2016} variance estimator and the bias-reduced variance estimator shown in Web Appendix A3 (suggested by \cite{Ozenne2021}) for the UNADJ estimator. Web Table S1 shows that the average estimated standard error calculated by \eqref{var_unadj} (AESE1), \eqref{unadjusted_var} (AESE2) and
the method developed in Web Appendix A3 (AESE3) are close to the Monte Carlo standard deviation (MCSD) for all estimators when sample size $n\geq 200$, and they are associated with nominal coverage, especially when the sample size is $n=400$. Furthermore, AESE1 and AESE3 are generally close, which is expected since these two corresponding variance estimators share the same underlying mechanism (projection theorem of U-statistics). However, with a small sample size (e.g., $n \leq 50$), coverage probabilities associated with AESE1 and AESE3 are closer to nominal than that associated with AESE2, especially under an unbalanced design. This result confirms that, if no covariate adjustment is performed, all three variance estimators are reliable when sample size is relatively moderate (e.g., $n\geq 200$), but only \eqref{var_unadj} and bias-reduced variance estimator in Web Appendix A3 are recommended for smaller sample sizes.

\subsubsection{Relative efficiency}
\begin{figure}
\centering
	\begin{minipage}[t]{3in}
		\centering
		\textbf{WR – (A)}          
		\includegraphics[width=2.5in]{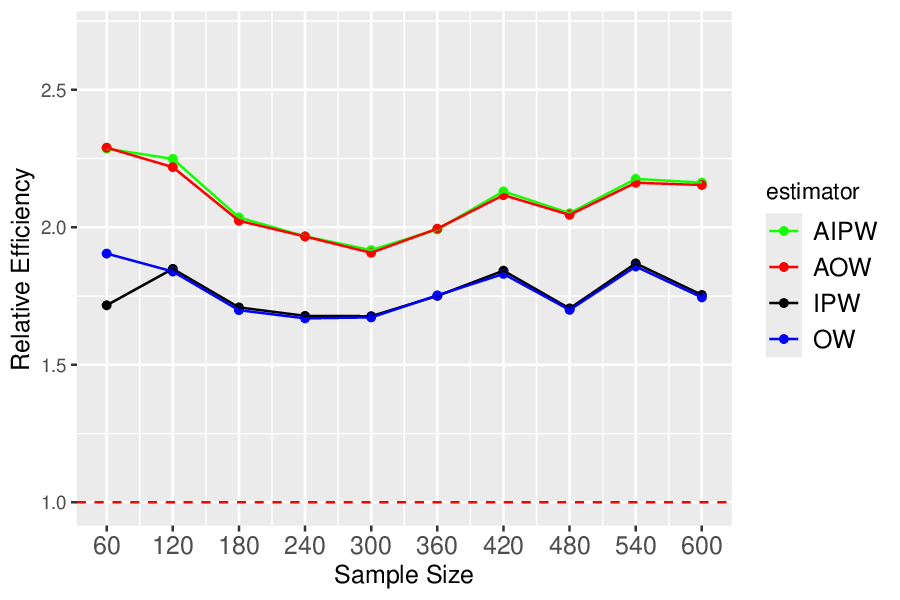}
	\end{minipage}%
	\begin{minipage}[t]{3in}
		\centering
		\textbf{WD – (A)}          
		\includegraphics[width=2.5in]{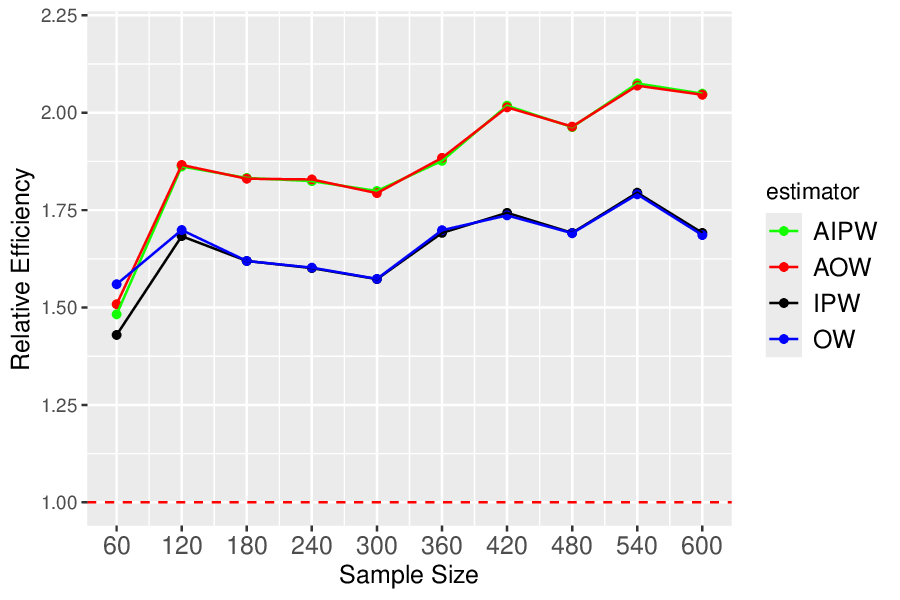}
	\end{minipage}%
	\vspace{0.05in}
	\begin{minipage}[t]{3in}
		\centering
		\textbf{WR – (B)}          
		\includegraphics[width=2.5in]{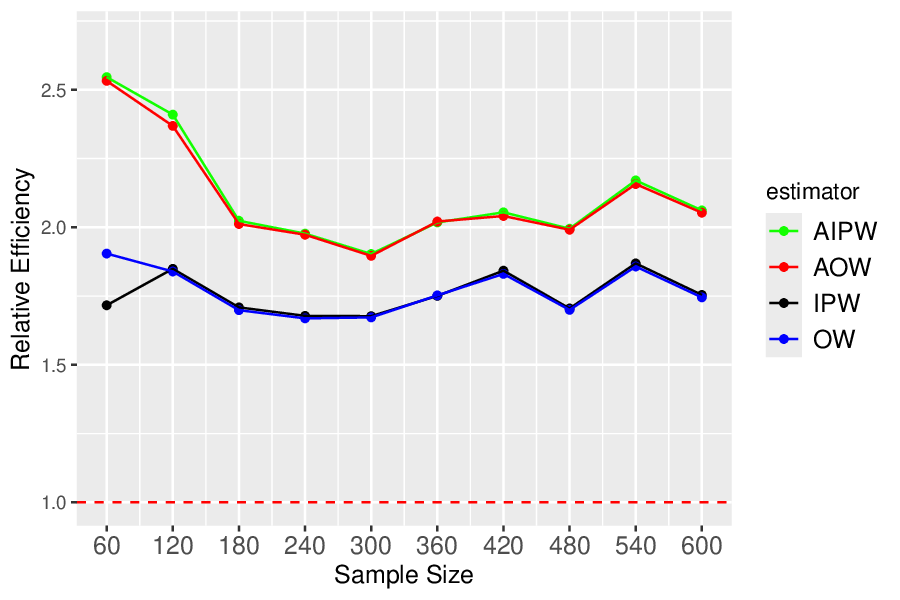}
	\end{minipage}%
	\begin{minipage}[t]{3in}
		\centering
		\textbf{WD – (B)}          
		\includegraphics[width=2.5in]{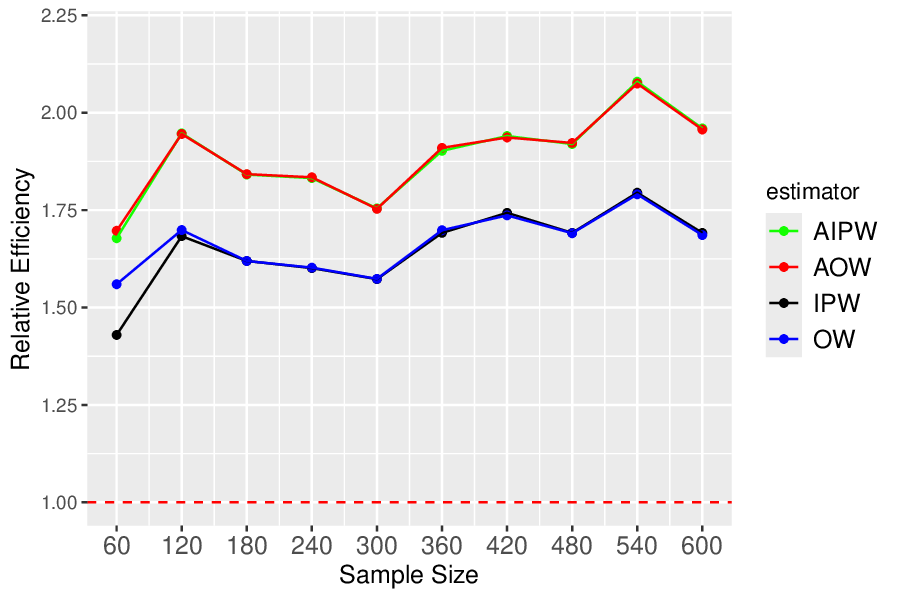}
	\end{minipage}%
	\vspace{0.05in}
	\begin{minipage}[t]{3in}
		\centering
		\textbf{WR – (C)}          
		\includegraphics[width=2.5in]{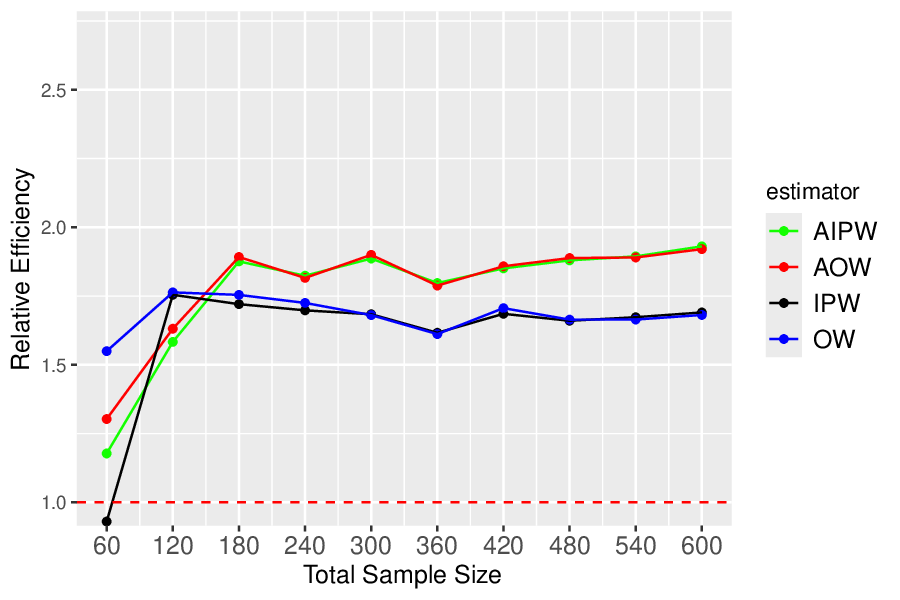}
	\end{minipage}%
	\begin{minipage}[t]{3in}
		\centering
		\textbf{WD – (C)}          
		\includegraphics[width=2.5in]{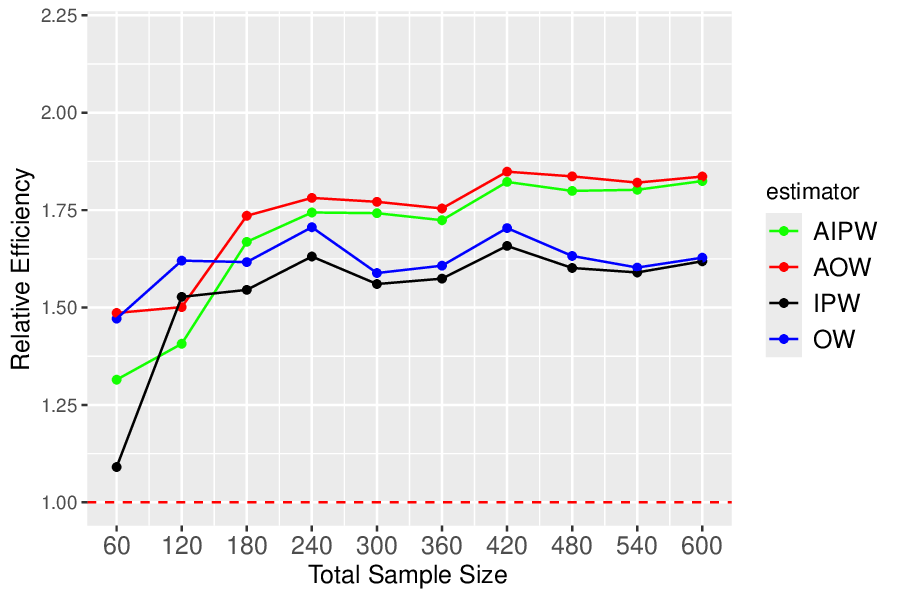}
	\end{minipage}%
	\vspace{0.05in}
	\begin{minipage}[t]{3in}
		\centering
		\textbf{WR – (D)}          
		\includegraphics[width=2.5in]{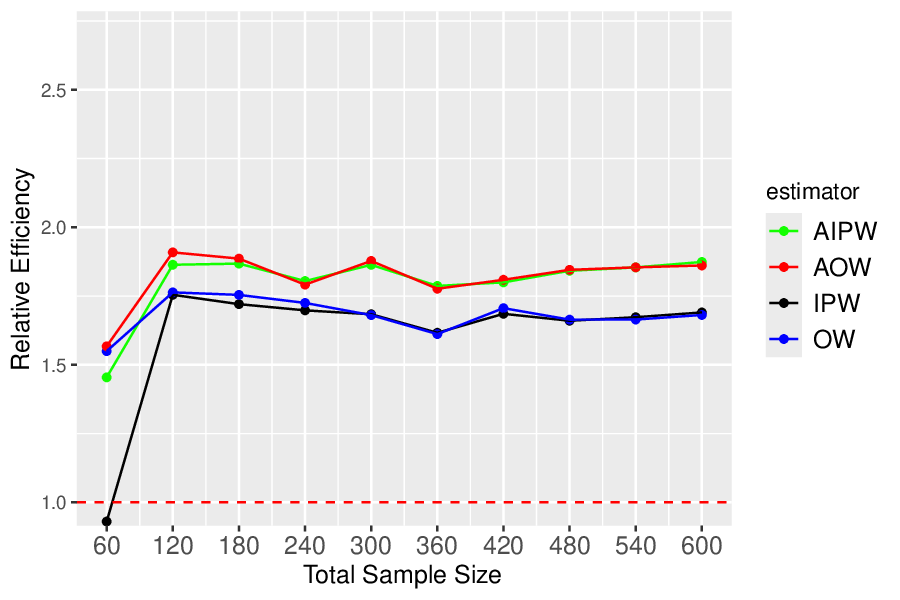}
	\end{minipage}
	\begin{minipage}[t]{3in}
		\centering
		\textbf{WD – (D)}          
		\includegraphics[width=2.5in]{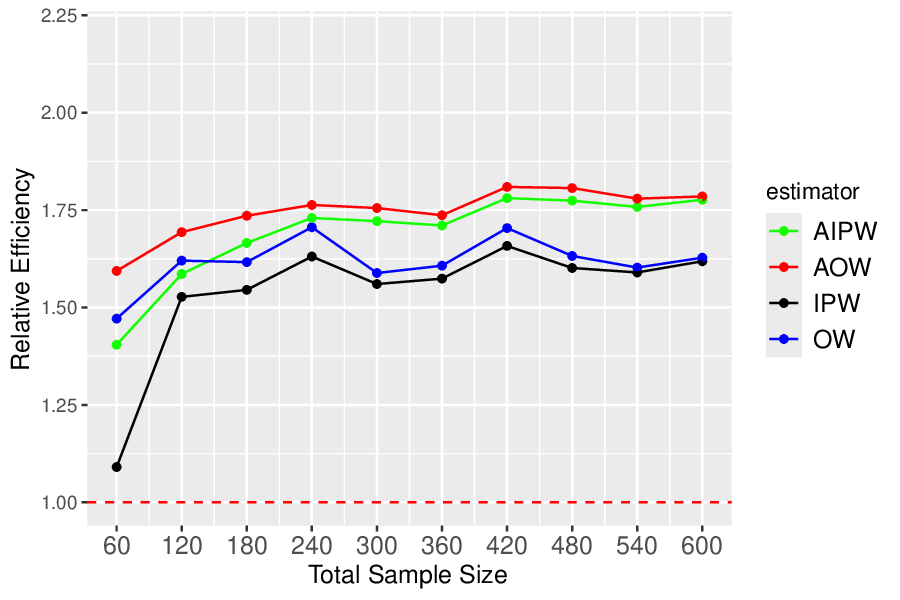}
	\end{minipage}
\caption{The relative efficiency of covariate-adjusted estimators for win ratio (WR) and  win difference (WD) when the data generating process follows the quadratic association model \eqref{quadratic}. Row (A): balanced design, correctly specified outcome regression model; Row (B) balanced design, mis‐specified outcome regression model; Row (C) unbalanced design, correctly specified outcome regression model; Row (D) unbalanced design, mis‐specified outcome regression model.}
\label{fig:quadratic_all_in_one}
\end{figure}

\begin{figure}
	\centering
	\begin{minipage}[t]{3in}
		\centering
		\textbf{WR – (A)}          
		\includegraphics[width=2.5in]{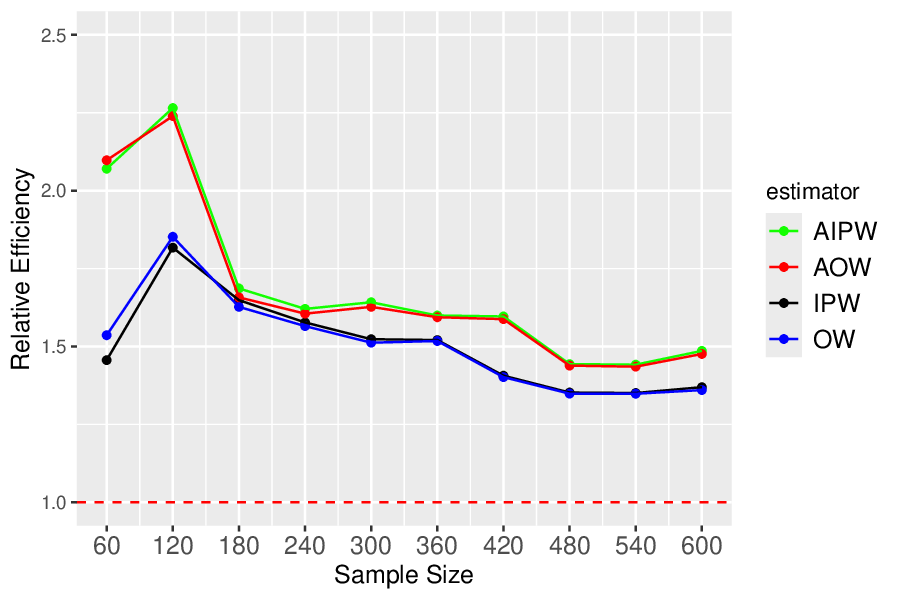}
	\end{minipage}%
	\begin{minipage}[t]{3in}
		\centering
		\textbf{WD – (A)}          
		\includegraphics[width=2.5in]{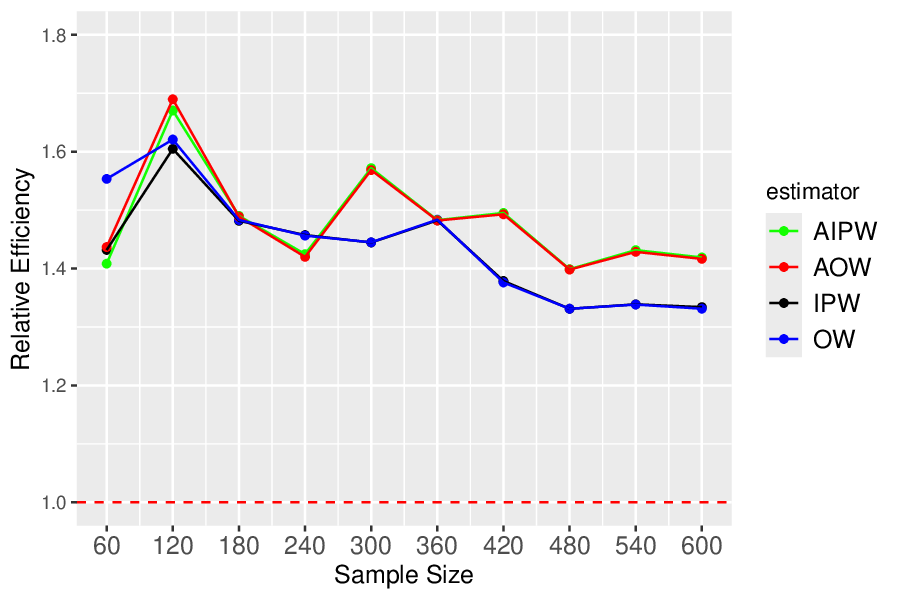}
	\end{minipage}%
	\vspace{0.05in}
	\begin{minipage}[t]{3in}
		\centering
		\textbf{WR – (B)}          
		\includegraphics[width=2.5in]{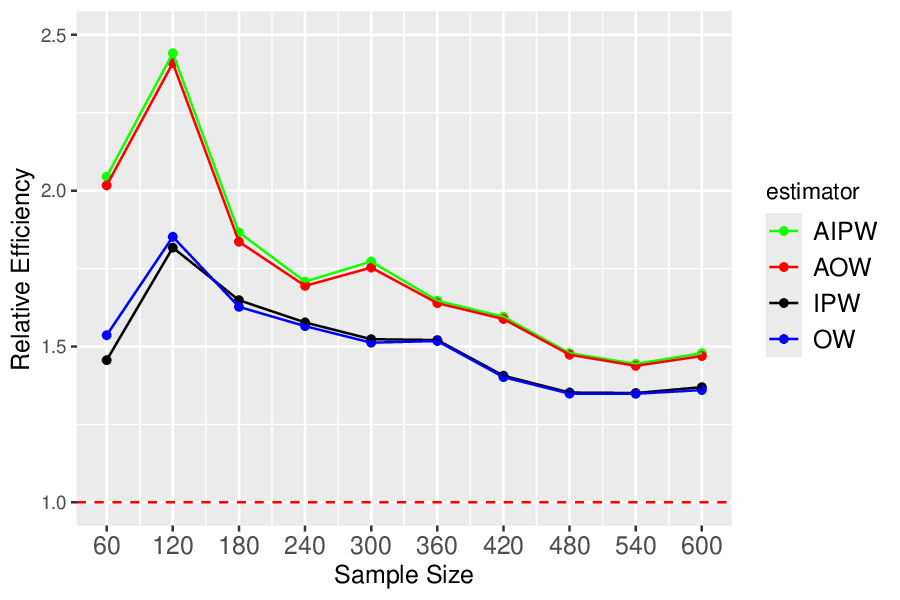}
	\end{minipage}%
	\begin{minipage}[t]{3in}
		\centering
		\textbf{WD – (B)}          
		\includegraphics[width=2.5in]{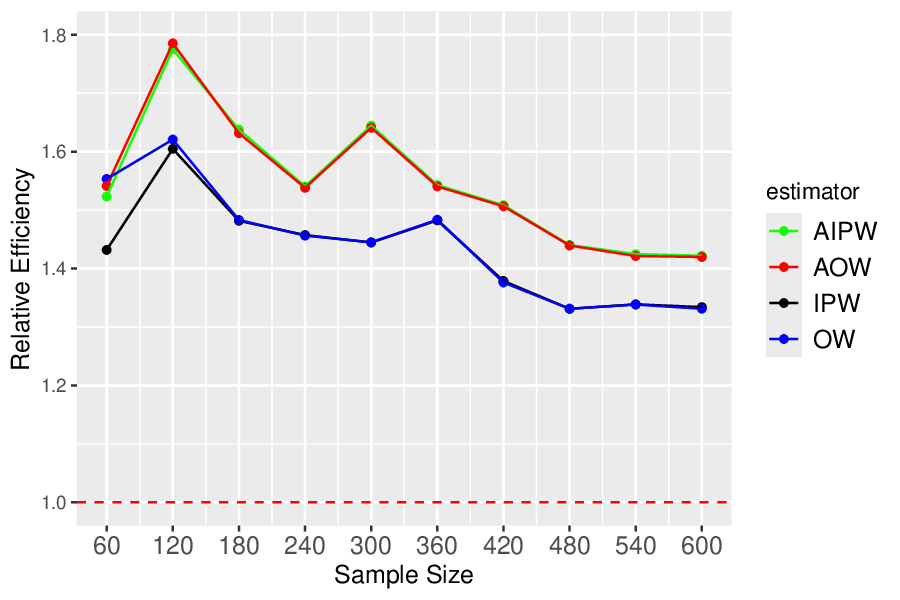}
	\end{minipage}%
	\vspace{0.05in}
	\begin{minipage}[t]{3in}
		\centering
		\textbf{WR – (C)}          
		\includegraphics[width=2.5in]{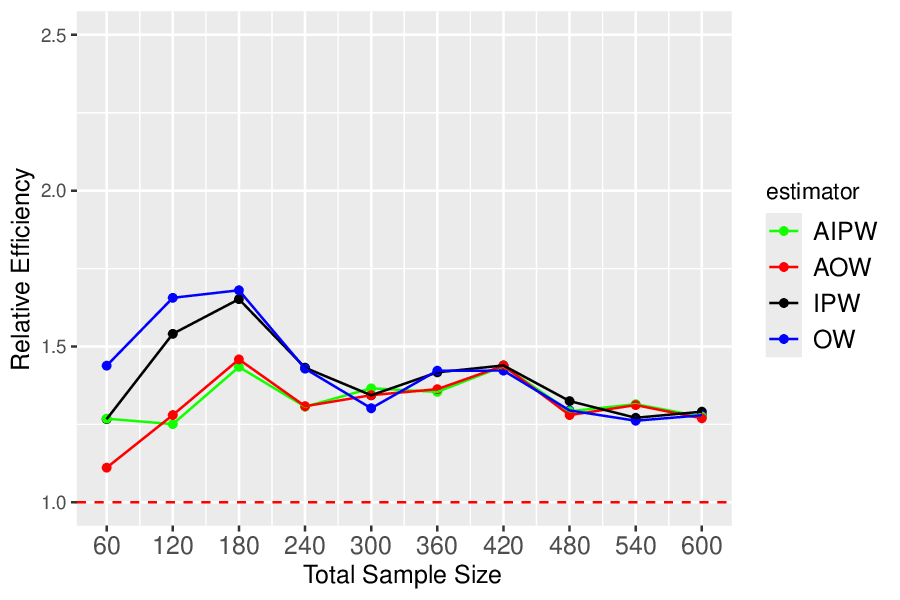}
	\end{minipage}%
	\begin{minipage}[t]{3in}
		\centering
		\textbf{WD – (C)}          
		\includegraphics[width=2.5in]{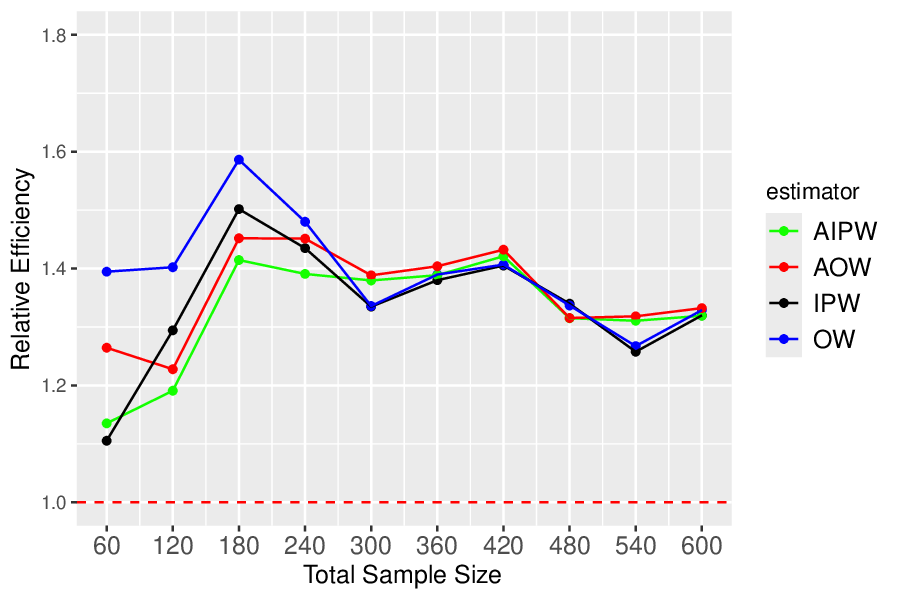}
	\end{minipage}%
	\vspace{0.05in}
	\begin{minipage}[t]{3in}
		\centering
		\textbf{WR – (D)}          
		\includegraphics[width=2.5in]{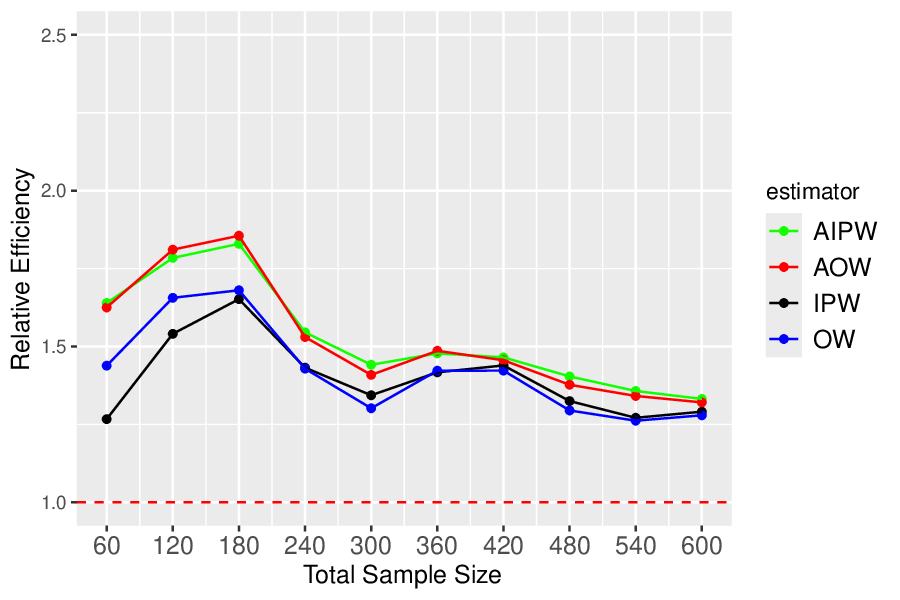}
	\end{minipage}
	\begin{minipage}[t]{3in}
		\centering
		\textbf{WD – (D)}          
		\includegraphics[width=2.5in]{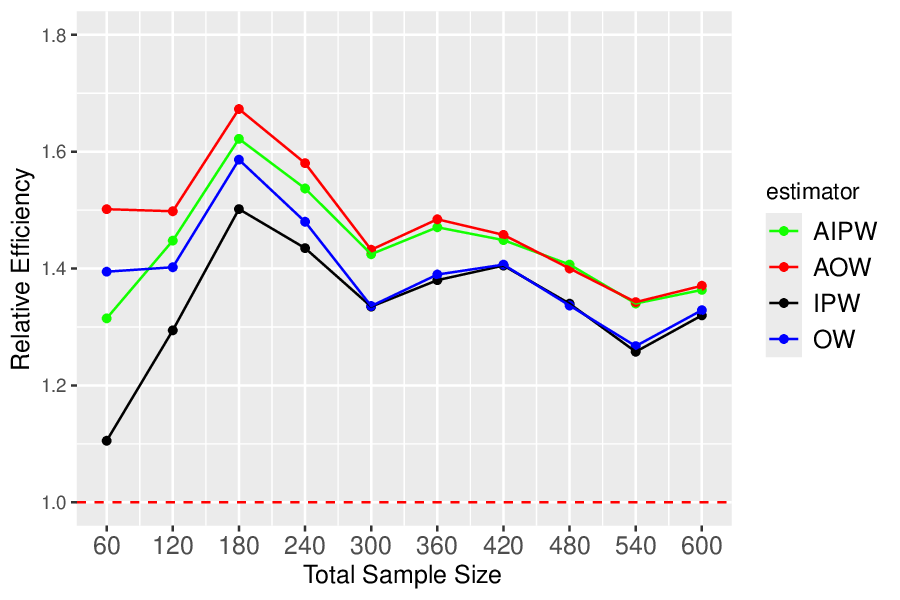}
	\end{minipage}
\caption{The relative efficiency of covariate-adjusted estimators for win ratio (WR) and  win difference (WD) when the data generating process follows the covariate interaction model \eqref{interaction}. Row (A): balanced design, correctly specified outcome regression model; Row (B) balanced design, mis‐specified outcome regression model; Row (C) unbalanced design, correctly specified outcome regression model; Row (D) unbalanced design, mis‐specified outcome regression model.}
\label{fig:interaction_all_in_one}
\end{figure}

Figures \ref{fig:quadratic_all_in_one}-\ref{fig:interaction_all_in_one} present the relative efficiency of IPW, OW, AIPW and AOW (AIPW-mis and AOW-mis) for WR and WD across sample sizes ranging from 60 to 600 when the data generating process follows the quadratic association model (Figure \ref{fig:quadratic_all_in_one}) and the covariate interaction model (Figure \ref{fig:interaction_all_in_one}), respectively. In each figure, we consider four scenarios: (A) balanced design with a correctly specified outcome regression model; (B) balanced design with a mis-specified outcome regression model; (C) unbalanced design with a correctly specified outcome regression model; (D) unbalanced design with a mis-specified outcome regression model.
\begin{center}
	[Figure \ref{fig:quadratic_all_in_one} about here.]
\end{center}
\begin{center}
	[Figure \ref{fig:interaction_all_in_one} about here.]
\end{center}

First, the propensity score estimators are notably more efficient than the unadjusted estimator, which confirms that adjusting for prognostic covariates can significantly improve the efficiency for estimating WR and WD, sometimes by as much as 2.5-fold. Furthermore, the augmented weighting estimators coupled with an additional outcome regression model (correctly or incorrectly specified) generally show additional efficiency gains across all scenarios. The only exceptions are (i) when the sample size is relatively small ($n\leq 120$) under an unbalanced design (row C in Figure \ref{fig:quadratic_all_in_one}), and (ii) when the data generating process follows the covariate interaction model under an unbalanced design (row C in Figure \ref{fig:interaction_all_in_one}). The likely reason for the latter case is that, in the true covariate interaction model \eqref{interaction}, there are 15 interaction terms in the true outcome model, posing stronger sample size requirements for the correctly specified AIPW and AOW estimators to gain further efficiency over the simpler propensity score weighting estimators. Except for that challenging scenario, when mis-specification of the outcome regression is present, the relative efficiency of AIPW and AOW is not much different than the case where the outcome regression is correctly specified, especially when the sample size is relative large, say $n\geq 420$ (see panel A vs panel B, panel C vs panel D in Figures \ref{fig:quadratic_all_in_one}-\ref{fig:interaction_all_in_one}). 
Finally, when mis-specification occurs in the scenario of an unbalanced design and the covariate interaction data generating model, the augmented weighting estimators still lead to higher efficiency over the propensity score weighting estimators (see panel D in Figure \ref{fig:interaction_all_in_one}) over the entire range of sample sizes. This suggests that, in relatively small RCTs with unbalanced assignment, careful consideration should be given to outcome model specification when using AIPW or AOW estimators, particularly regarding the number of covariates to adjust for, due to the tradeoff between model complexity and limited sample size.

These simulation results also shed light on the efficiency comparison between using IPW and OW for covariate adjustment when estimating WR and WD. Under a balanced design, we find that IPW and OW, as well as AIPW and AOW often perform similarly in terms of efficiency, especially as the sample size increases. The use of overlap weights can have a slight efficiency advantage when the sample size is small, say, no more than $120$. The efficiency difference between the use of different weights (IPW versus OW, AIPW versus OW) is more obvious under an unbalanced design. Under an unbalanced design which inherently offers fewer pairwise comparisons, OW-based estimators often demonstrate superior efficiency over their IPW-based counterparts, especially when sample sizes are smaller. Furthermore, the efficiency advantages of AOW over AIPW and OW over IPW are more pronounced when estimating the WD across all scenarios within unbalanced designs and when estimating the WR in scenarios under the quadratic association data generating model within relatively small sample size (e.g., $n\leq 120$) unbalanced design. 
Thus, the relative efficiency of OW and AOW estimators is most beneficial in smaller studies or those with unbalanced randomization.

\subsection{Additional simulations for the testing procedures}
Theorem \ref{prop1} and \ref{prop2} also provide a principled basis for deriving testing procedures for $\Delta\in\{\log(\theta),\eta\}$. Consider the Wald-type test based on the studentized statistic $Z_{*}=\widehat{\Delta}/\widehat{\sigma}_{*}$, where $\widehat{\sigma}_{*}$ consistently estimates the true standard deviation $\sigma_{*}$. Then under the null hypothesis $H_0:\Delta=0$, we have $Z_{*}\stackrel{d}\sim N(0,1)$. 
Note that for both model \eqref{quadratic} and model \eqref{interaction}, if we let the regression coefficient $\delta=0$, $Y_i(1)$ and $Y_j(0)$ are generated from the exact same distribution, allowing us to explore type I error of UNADJ, IPW, OW, AIPW, AOW, AIPW-mis and AOW-mis. The empirical type I error rate results under Wald $z$ test are reported in Table \ref{tb:em_type_one}.  We observe that under the null, the empirical type I error rates from UNADJ, IPW, OW, AOW, AIPW-mis and AOW-mis generally remain close to the nominal 5\% level, but the empirical type I error rate from AIPW can be inflated and this is more pronounced for a small sample and an unbalanced design. This result is consistent with the simulation findings in Section \ref{mc_var} and is likely due to the complexity of the true outcome model. That is, when the outcome regression includes many interaction terms relative to the available sample size, estimation of the augmentation component and its associated variance can become less stable, leading to mild finite-sample inflation of the Wald test.

\begin{center}
	[Table \ref{tb:em_type_one} about here.]
\end{center}

Further, we conducted additional simulations to evaluate whether results in Section \ref{mc_bias} and \ref{mc_var} will be affected by strong correlation among covariates or not. Detailed data generation when covariates are strongly correlated is presented in Web Appendix E, while corresponding point estimates, variance and interval estimates, and empirical type I error rate are shown in Tables S2, S3 and S4, respectively. Since results in Table S2 and Table 1, Table S3 and Table 2, as well as Table S4 and Table 3, are similar and comparable, it is concluded that our findings based on Section \ref{subsec:simresults} are not affected by strong correlation among covariates.

\section{An Illustrative Application to the ORCHID clinical trial}\label{sec:ORCHID}
To illustrate our proposed methods, we reanalyze a completed RCT --- the Outcomes Related to COVID-19 Treated With Hydroxychloroquine Among Inpatients With Symptomatic Disease (ORCHID). This multi-center, double-blind, placebo-controlled trial was carried out under the PETAL Clinical Trials Network between April and June 2020, and aims to test the hypothesis that, compared with placebo, hydroxychloroquine improves clinical outcomes for adults hospitalized with COVID-19 \citep{self2020effect}. The study enrolled 479 adults hospitalized with confirmed SARS-CoV-2 infection and respiratory symptoms, and randomly assigned them to receive either hydroxychloroquine or placebo. The primary outcome was the patient's clinical status at 14 days, assessed via a 7-point ordinal scale: (1) Not hospitalized, performing normal activities; (2) Not hospitalized, but unable to perform normal activities; (3) Hospitalized, not requiring supplemental oxygen; (4) Hospitalized, requiring supplemental oxygen via nasal cannula; (5) Hospitalized, requiring respiratory support via non-rebreather mask, or high-flow nasal cannula oxygen therapy, or noninvasive mechanical ventilation; (6) Hospitalized, requiring extracorporeal membrane oxygen, invasive mechanical ventilation, or both; (7) Death; all of which are based on the World Health Organization-recommended 7-point COVID Outcomes Scale \citep{Cummings2020}.
The trial also included 12 secondary outcomes, including 7-day and 28-day mortality outcomes.
Despite early optimism based on preclinical antiviral and immunomodulatory properties of hydroxychloroquine, the primary analysis based on a multivariable proportional odds model obtained an adjusted odds ratio (aOR, 
	a value greater than 1.0 indicating more favorable outcomes with hydroxychloroquine than placebo) of 1.02, with 95\% confidence interval [0.73, 1.42]. This result means that, compared with placebo, treatment with hydroxychloroquine did not significantly improve clinical status at day 14. We note that this estimand is defined through the regression model rather than as a marginal contrast of potential outcomes, and that it relies on the proportional odds assumption; in contrast, the win estimands we will consider below are marginal, model-free summaries based on pairwise comparisons of potential outcomes, providing a complementary perspective on the same data.

In this data example, we leverage our win statistics framework to analyze the ordinal clinical status outcomes measured at 7, 14, and 28 days after randomization, with the 14-day outcome corresponding to the primary outcome in the original ORCHID trial analysis. We adjust for a comprehensive set of potential prognostic factors, including body mass index, ICU status at baseline, standardized age, race, sex, baseline oxygen saturation, respiratory rate, and the Glasgow Coma Scale component of the SOFA score. These covariates are adjusted for as linear terms in the logistic propensity score model for IPW and OW estimators. Moreover, for augmented weighting estimators, we also adjust for the same set of covariates as linear terms in the ordinal logistic outcome model $\mu(\cdot,\cdot;\alpha)$. To quantify the treatment effect of hydroxychloroquine compared to placebo, we estimate the WR and WD with the five different methods (UNADJ, IPW, OW, AIPW and AOW). The point estimate, standard error (SE) calculated by corresponding closed-form variance estimation (\eqref{var_unadj}, (A.8), (A.9), (A.13) and (A.14) in Web Appendix, respectively) as well as the 95\% confidence intervals are summarized in Table \ref{tab:analysis1}. In addition, to measure the finite-sample variance reduction from weighting and augmented weighting estimators (compared to the unadjusted estimator) through covariate adjustment, we compute the proportion of variance reduction compared to the unadjusted estimator (PVR), defined as $\text{PVR}={(\text{SE}^2_{\text{UNADJ}}-\text{SE}^2_{\text{Estimator}})}/{\text{SE}^2_{\text{UNADJ}}}\times 100\%$,  
where $\text{Estimator} \in\{\text{IPW, OW, AIPW, AOW}\}$.

\begin{center}
	[Table \ref{tab:analysis1} about here.]
\end{center} 

According to Table \ref{tab:analysis1}, we observe that across all three time points (Day 7, Day 14, and Day 28), the estimated win ratios for hydroxychloroquine versus placebo from the five proposed methods carry magnitudes larger than 1, with 95\% confidence intervals consistently including the null, indicating no statistically significant treatment effect on the pairwise-comparison scale. The direction and statistical conclusions when the win difference is reported are the same. Both estimated WR and WD conclusions are consistent with the original analysis in \citet{self2020effect}. To interpret the results, we consider the AOW estimates for the outcome at Day 7 as an example. Under this approach, the WR comparing hydroxychloroquine to placebo was 1.066 (95\% CI: 0.855–1.278), indicating that the probability of a more favorable outcome for a randomly selected hydroxychloroquine-treated patient compared to a placebo patient was 6.6\% higher than the reverse. Similarly, the WD comparing hydroxychloroquine to placebo was 0.026 (95\% CI: -0.055–0.108), indicating that the probability of a more favorable outcome for a randomly selected hydroxychloroquine-treated patient exceeded that for a placebo patient by 2.6 percentage points. 
Finally, comparing across methods, the point estimates are generally similar regardless of whether baseline covariate adjustment is used. However, the standard error estimates after covariate adjustment are typically smaller. For example, when estimating the WR for the Day 14 outcome, covariate-adjusted weighting estimators (IPW and OW) and augmented weighting estimators (AIPW and AOW) achieve over 30\% variance reduction (34.4\%, 33.3\%, 34.1\% and 33.5\%, respectively) compared with unadjusted estimators. Overall, we find that the amount of variance reduction is relatively consistent across the weighting and augmented weighting estimators, suggesting that including an additional ordinal outcome regression beyond weighting, in this example, does not further reduce the variance of the WR and WD estimators.

\section{Discussion}\label{sec:Discussion}
In this article, we have introduced covariate-adjustment methods to estimate the win ratio and win difference. We first developed propensity score weighting estimators --- in particular, IPW and OW --- for win statistics. These estimators extend the prior work of \citet{zeng2021propensity} from the conventional average treatment effect estimand to a wider class of pairwise comparison estimands, for which win ratios and win differences are important special cases. Second, inspired by the form of the augmented weighting estimator in \citet{Mao2019} for analyzing observational data, we further introduced a class of augmented weighting estimators for win estimands that leveraged an additional outcome regression (in addition to weighting) to potentially further improve efficiency. Both weighting and augmented weighting estimators (IPW or OW) were constructed using stabilized weights such that they are H{\'a}jek-type estimators, different from the unstabilized IPW and augmented inverse probability weighting estimators previously proposed for observational studies \citep{Mao2018}. We applied the U-statistics theory to derive the asymptotic variance of all unadjusted and covariate-adjusted estimators, and obtained the closed-form variance of the weighting and augmented weighting estimators from the sample variance of the estimated influence function. Furthermore, we showed that the influence function of the weighting estimators can be regarded as the influence function of the unadjusted estimator minus its projection to the nuisance tangent space, and proved that the asymptotic variance of the propensity score weighting estimators is no larger than that of the unadjusted estimator. This has important implications, as it provides a compelling rationale for covariate adjustment when targeting win estimands in RCTs, reinforcing the principle that baseline covariates should almost always be leveraged when available rather than ignored. More broadly, this finding strengthens the position reflected in the FDA guidance for COVID-19 trials \citep{FDA2020COVID}, which stated: ``\emph{To improve the precision of treatment effect estimation and inference, sponsors should consider adjusting for prespecified prognostic baseline covariates \ldots and should propose methods of covariate adjustment}.'' Our work thus illustrates the operational steps involved in estimating win estimands with the unadjusted and covariate-adjusted win-statistic estimators in the \texttt{winPSW}
R package.

Through extensive simulations, we found that, compared to the unadjusted estimator, propensity score weighting estimators are clearly more statistically efficient by leveraging prognostic baseline covariates for estimating both the WR and WD. Furthermore, the augmented weighting estimators showed a further improvement in efficiency, except under unbalanced randomization and with correctly-specified outcome regression based on the interaction model \eqref{interaction}; in that case, the performance of the propensity score weighting and augmented weighting estimators performed rather similarly, even when the sample size increases. When the sample size was relatively small, e.g., $n\leq 120$, the OW estimator is marginally more efficient than the IPW estimator. Under an unbalanced design with smaller sample sizes, AOW is also more efficient than AIPW and empirical type I error rate of AOW is closer to the nominal 5\% level. But in most cases with sufficient sample sizes, our observations are that OW and IPW, AOW and AIPW have similar performance when addressing the WR and WD, which is consistent with prior findings for comparing IPW and OW in randomized trials with a binary outcome \citep{zeng2021propensity}. Finally, as a cautionary note, we observed that when estimating the WR, the IPW estimator could be less efficient than the unadjusted estimator when the sample size was extremely small (e.g., $n=60$) and the degree of treatment effect heterogeneity was moderate. Based on this observation, we advocate using the OW estimator when the sample size is limited, to safeguard potential finite-sample efficiency loss. 

Two estimands, the WR and WD, served as our running examples for demonstrating efficiency gains from propensity score weighting and augmented weighting. We caution, however, that the WR alone may be a misleading summary of an ordinal treatment effect in the presence of ties, and this limitation has been the subject of active debate \citep{Ajufo2023,Butler2024,Davison2025}. This is because WR is the ratio of winning to losing comparisons, and is invariant to the absolute number of decisive comparisons and therefore does not always accurately reflect the magnitude of evidence. For example, a trial with 3 wins, 1 loss, and 4996 ties and a trial with 3000 wins, 1000 losses, and 1000 ties both yield a WR of 3, even though the latter conveys a far stronger treatment benefit. The WD does not share this insensitivity, as it normalizes by all pairwise comparisons rather than only decisive ones, and would distinguish the two trials above (0.0004 versus 0.40). For this reason, we recommend reporting the WR and WD together when analyzing ordinal outcomes, rather than relying on the WR alone, so that both the relative and absolute views of the treatment effect are conveyed. Alternatively, a useful measure that incorporates ties is the win odds (WO, or success odds) \citep{Dong2020wo,brunner2021win}, defined as 
	\begin{eqnarray*}\label{win_odd}
		\zeta = \frac{\mathbb{P}\left\{Y_i(1)>Y_j(0)\right\}+\frac{1}{2}\mathbb{P}\left\{Y_i(1)=Y_j(0)\right\}}{\mathbb{P}\left\{Y_i(1)<Y_j(0)\right\}+\frac{1}{2}\mathbb{P}\left\{Y_i(1)=Y_j(0)\right\}}.
	\end{eqnarray*}
	After incorporating tied comparisons into the analysis, the WO not only provides a more accurate reflection of similarity between groups and ensures greater numerical stability, but also improves interpretation since it can be interpreted as a true odds \citep{Verbeeck2023}. In principle, our covariate-adjusted estimators can be generalized to accommodate WO. To proceed, the win probability defined in \eqref{win_prob} should first be modified to be the probabilistic index, that is, $\tau_{1}=\mathbb{P}\left\{Y_i(1)>Y_j(0)\right\}+\mathbb{P}\left\{Y_i(1)=Y_j(0)\right\}/2$. Second, when augmented weighting estimators are used, the outcome model $\mu(\cdot,\cdot;\alpha)$ could be fitted by the more general, probabilistic index model (PIM) proposed by \cite{Thas2012}. Then, by using the same procedure for weighting and augmented weighting estimators described in Section \ref{sec:weighting} and Section \ref{sec:augmented}, we can obtain corresponding WO estimators. We expect similar patterns for the covariate-adjusted WO estimator but defer a full development to future research. 
	Finally, \cite{Harrell2021} proposed a Markov proportional odds ordinal logistic model that analyzes longitudinal ordinal outcomes conditional on baseline covariates (including treatment) and the previous state, allowing the treatment effect to be estimated as a function of time. There are, however, two key distinctions from our work. Above all, their approach requires the ordinal outcome to be measured repeatedly over time for each individual, whereas our estimators target a single cross-sectional ordinal outcome. In addition, the native parameter of the transition model is a within-patient odds ratio conditional on the previous state, which, even after marginalization, yields time-indexed state-occupancy probabilities rather than the marginal, between-patient pairwise-comparison estimands considered in our framework. Although their model can provide more efficient derived estimates than simple patient-level summary statistics under a correctly specified transition model, the differing estimands and data structures mean that a direct efficiency comparison is not straightforward, and we leave a formal investigation to future research.

We conclude by summarizing two promising future research directions. First, although our study focuses on the weighting and augmented weighting estimators in the context of randomized trials, the proposed estimators and software are equally applicable to observational studies with ignorable treatment assignment, though the target estimand based on pairwise comparisons will differ depending on the choice of the tilting function $h(X_i,X_j)$; see \citet{Li2018} for a detailed elaboration in the context of the weighted average treatment effect estimand (not constructed based on pairwise comparisons). For example, although IPW targets the entire combined population from treatment and control, OW would target the subpopulation with sufficient covariate overlap and clinical equipoise \citep{Li2018}. By shifting the goalpost, OW has theoretically and empirically proven to be the most efficient choice among the class of balancing weights \citep{Li2018,Li2019,cao2024using}, and is the recommended approach to address lack of positivity in the treatment assignment \citep{Li2019address}. However, the optimality of OW has so far been established only for estimating the average treatment effect; additional work is needed to establish its optimality for estimating pairwise comparison estimands in the form of \eqref{balanced_ate}. Second, although we have focused on a univariate ordinal outcome, the win-statistics framework is most often applied to prioritized composite endpoints \citep{buyse2025}, and our estimators can be extended to this setting. That is, as long as the win and loss functions are well defined, the machinery of Sections \ref{sec:weighting} and \ref{sec:augmented} carries over directly, with the augmented estimators additionally requiring a plausible working outcome model (e.g., through the PIM). A key complication arises when the prioritized components involve right-censored time-to-event outcomes, where the win and loss estimands are defined in the absence of censoring. In such cases, the inverse probability of censoring weighting has been suggested in the literature \citep{dong2020ipcw,Cao2026,fang2026improving}, and a full development of covariate-adjusted win statistics for censored composite endpoints is left to future research.

\section*{Acknowledgement}
F. Li, G. Tong and PJ Heagerty are supported by the United States National Institutes of Health (NIH), National Heart, Lung, and Blood Institute (NHLBI, grant number 1R01HL178513). K. Davis-Plourde is supported by CTSA Grant Number UL1 TR001863 from the National Center for Advancing Translational Science (NCATS), a component of the National Institutes of Health (NIH). All statements in this report, including its findings and conclusions, are solely those of the authors and do not necessarily represent the views of the NIH. 

\section*{Data Availability Statement}
The de-identified data from the ORCHID study are publicly available from the NHLBI BioLINCC website and can be requested at \url{https://biolincc.nhlbi.nih.gov/studies/petal_orchid/}. F. Li obtained access to the publicly available ORCHID study data according to a Data Use Agreement from BioLINCC and performed the data analysis in Section 6.

\section*{Supporting information}
R code used in this article for simulation studies can be found at \url{https://github.com/Zhiqiangcao/winPSW/tree/main/simulation_code}.
Additional supporting information including Web Appendices A--F can be found at Github platform \url{https://github.com/Zhiqiangcao/winPSW}.

\bibliographystyle{jasa3}
\bibliography{DRGEN}

\clearpage

\clearpage

\clearpage
\begin{table}
\footnotesize
\begin{center}
\caption{Relative bias (\%) of WR and WD \textcolor{blue}{estimators} under balanced and unbalanced designs. UNADJ: the unadjusted estimator; IPW: inverse probability weighting; OW: overlap weighting; AIPW: augmented inverse probability weighting; AOW: augmented overlap weighting; AIPW-mis: augmented inverse probability weighting with a misspecified working outcome model; and AOW-mis: augmented overlap weighting with a misspecified working outcome model.}
		\label{tb:RB}
		\begin{adjustbox}{width=1\textwidth}
			\begin{tabular}{llccrrrrrrrr}
				\hline
				True model &Design & $n$ & Estimand & Truth & UNADJ & IPW & OW & AIPW & AOW & AIPW-mis & AOW-mis   \\
				\hline
				Quadratic & Balanced &$200$ &WR &1.453 & 4.73 & 1.84 & 2.05 & 1.82 & 2.02 & 2.10 & 2.31 \\ 
				&          &      &WD &0.091 &1.65 & -0.74 & -0.31 & 0.36 & 0.76 & 0.87 & 1.29 \\ 
				&          &$300$ &WR &1.453 & 1.17 & 0.90 & 1.00 & 0.57 & 0.70 & 0.42 & 0.56 \\ 
				&          &      &WD &0.091 &-3.05 &-0.51 &-0.31 &-0.82 &-0.54 &-1.20 &-0.91 \\ 
				&          &$400$ &WR &1.453 &2.58 & 0.88 & 1.00 & 0.14 & 0.24 & 0.12 & 0.22 \\  
				&          &      &WD &0.091 &2.19 & 0.17 & 0.44 &-1.05 &-0.81 &-1.25 &-1.02 \\  
				\\
				&Unbalanced&$200$ &WR &1.450 & 4.19 & 2.92 & 3.20 & 2.48 & 2.88 & 2.23 & 2.52 \\ 
				&          &      &WD &0.091 & 0.72 & 1.32 & 1.33 & 0.85 & 1.42 & 0.60 & 0.74 \\  
				&          &$300$ &WR &1.450 &3.01 & 1.36 & 1.59 & 1.22 & 1.47 & 1.11 & 1.35 \\  
				&          &      &WD &0.091 &2.38 & 0.47 & 0.89 & 0.68 & 1.14 & 0.38 & 0.77 \\  
				&          &$400$ &WR &1.450 &0.10 &-0.10 & 0.20 & 0.30 & 0.55 & 0.28 & 0.50 \\ 
				&          &      &WD &0.091 &-3.96&-2.02 &-1.53 &-0.55 &-0.17 &-0.61 & -0.29 \\ 
				\\
				Interaction& Balanced &$200$ &WR &1.663 &4.86 & 1.86 & 2.13 & 1.77 & 2.07 & 2.26 & 2.54 \\ 
				&          &      &WD &0.126 &0.77 &-0.71 &-0.38 &-1.20 &-0.82 & 0.48 & 0.82 \\  
				&          &$300$ &WR &1.663 &1.68 & 1.37 & 1.53 & 0.52 & 0.71 & 1.17 & 1.28 \\ 
				&          &      &WD &0.126 &-1.46& 0.30 & 0.52 &-1.18 & -0.88 & 0.40 & 0.49 \\  
				&          &$400$ &WR &1.663 &2.35 & 0.58 & 0.75 &-0.09 & 0.05 & -0.04 & 0.10 \\   
				&          &      &WD &0.126 &0.92 &-0.45 &-0.18 &-1.46 &-1.23 & -1.24 & -1.02 \\  
				\\
				&Unbalanced&$200$ &WR &1.657 &3.85 & 2.07 & 2.40 & 3.69 & 3.99 & 1.26 & 1.66 \\ 
				&          &      &WD &0.125 &-0.01&-0.22 &-0.22 & 1.03 & 1.29 &-1.37 &-1.08 \\   
				&          &$300$ &WR &1.657 &3.45 & 1.55 & 1.86 & 1.94 & 2.26 & 1.29 & 1.63 \\  
				&          &      &WD &0.125 &2.11 & 0.53 & 0.94 & 1.03 & 1.49 & 0.50 & 0.97 \\  
				&          &$400$ &WR &1.657 & 0.83 & 0.63 & 1.03 & 1.31 & 1.65 & 0.79 & 1.19 \\ 
				&          &      &WD &0.125 &-1.65 & -0.20 & 0.22 & 1.21 & 1.51 & 0.47 & 0.86 \\  
				\hline 
			\end{tabular}
		\end{adjustbox}
	\end{center}
\end{table}

\clearpage

\begin{table}
\footnotesize
\centering
\caption{The ratio between the average estimated variance over the true, empirical Monte Carlo variance ($\text{\{Est Var\}/\{MC Var\}}$), as well as the associated 95\% confidence interval coverage rate of different methods. UNADJ: the unadjusted estimator; IPW: inverse probability weighting; OW: overlap weighting; AIPW: augmented inverse probability weighting; AOW: augmented overlap weighting; AIPW-mis: augmented inverse probability weighting with a misspecified working outcome model; and AOW-mis: augmented overlap weighting with a misspecified working outcome model.}
	\label{tab:variance_and_coverage}
	\begin{adjustbox}{width=1\textwidth}
		\begin{tabular}{llccccccc|ccccccc}
			\toprule
			& & \multicolumn{7}{c}{$\text{\{Est Var\}/\{MC Var\}}$} & \multicolumn{7}{c}{95\% confidence interval coverage} \\
			\cmidrule(lr){3-9} \cmidrule(lr){10-16}
			True model & $n$ & UNADJ & IPW & OW & AIPW & AOW & AIPW-mis &AOW-mis & UNADJ & IPW & OW & AIPW & AOW &AIPW-mis &AOW-mis \\
			\midrule
			& &\multicolumn{14}{c}{WR under balanced design} \\
			\cmidrule(lr){3-16}
			Quadratic &200 &0.962 & 0.981 & 0.981 & 0.877 & 0.897 & 0.899 & 0.918 & 0.932 & 0.945 & 0.940                & 0.927 & 0.931 & 0.930 & 0.932 \\ 
			&300 & 0.984 & 1.020 & 1.016 & 0.897 & 0.905 & 0.939 & 0.948 & 0.944 & 0.950 &    
			0.949 & 0.942 & 0.945 & 0.936 & 0.939 \\ 
			&400 & 0.976 & 0.962 & 0.967 & 0.928 & 0.940 & 0.930 & 0.942 & 0.946 & 0.947 &   
			0.946 & 0.934 & 0.936 & 0.937 & 0.939 \\  
			& &\multicolumn{14}{c}{WD under balanced design} \\
			\cmidrule(lr){3-16}
			&200 &0.998 & 0.981 & 0.976 & 0.903 & 0.926 & 0.932 & 0.955 & 0.949 & 0.950 & 0.947      & 0.926 & 0.932 & 0.937 & 0.942 \\ 
			&300 & 1.038 & 1.060 & 1.058 & 0.932 & 0.943 & 0.975 & 0.987 & 0.952 & 0.953 &    
			0.955 & 0.937 & 0.937 & 0.944 & 0.944 \\ 
			&400 & 0.984 & 0.977 & 0.980 & 0.941 & 0.954 & 0.948 & 0.959 & 0.941 & 0.955 &  
			0.956 & 0.944 & 0.945 & 0.945 & 0.946 \\
			& &\multicolumn{14}{c}{WR under unbalanced design} \\
			\cmidrule(lr){3-16}
			&200&0.912 & 0.865 & 0.854 & 0.704 & 0.863 & 0.798 & 0.954 & 0.939 & 0.935 & 0.944  
			& 0.911 & 0.937 & 0.925 & 0.949 \\ 
			&300&1.025 & 0.988 & 0.995 & 0.896 & 1.065 & 0.941 & 1.106 & 0.950 & 0.942 & 0.949 
			& 0.933 & 0.957 & 0.951 & 0.961 \\ 
			&400&0.971 & 1.042 & 1.049 & 0.937 & 1.126 & 0.996 & 1.188 & 0.942 & 0.948 & 0.953 
			& 0.944 & 0.964 & 0.952 & 0.967 \\ 
			& &\multicolumn{14}{c}{WD under unbalanced design} \\
			\cmidrule(lr){3-16}
			&200&0.976 & 0.909 & 0.908 & 0.726 & 0.889 & 0.818 & 0.983 & 0.942 & 0.936 & 0.939 & 0.898 & 0.932 & 0.914 & 0.945 \\ 
			&300&1.046 & 1.013 & 1.020 & 0.908 & 1.066 & 0.957 & 1.116 & 0.955 & 0.945 & 0.951 & 0.937 & 0.957 & 0.944 & 0.961 \\ 
			&400& 0.975 & 1.054 & 1.062 & 0.937 & 1.113 & 0.994 & 1.172 & 0.951 & 0.956 & 0.952 & 0.945 & 0.964 & 0.953 & 0.964 \\
			\\
			& &\multicolumn{14}{c}{WR under balanced design} \\
			\cmidrule(lr){3-16}
			Interaction &200 &0.905 & 0.955 & 0.948 & 0.622 & 0.635 & 0.832 & 0.846 & 0.933 & 0.942 & 0.947 & 0.887 
			& 0.895 & 0.932 & 0.937 \\ 
			&300 & 0.968 & 0.929 & 0.928 & 0.702 & 0.714 & 0.867 & 0.920 & 0.943 & 0.944 & 0.945 & 0.902 & 0.905 & 0.936 & 0.931 \\ 
			&400 &0.971 & 1.014 & 1.020 & 0.842 & 0.857 & 0.946 & 0.959 & 0.948 & 0.951 & 0.951 & 0.916 & 0.919 & 0.933 & 0.936 \\  
			& &\multicolumn{14}{c}{WD under balanced design} \\
			\cmidrule(lr){3-16}
			&200 &0.999 & 0.989 & 0.980 & 0.680 & 0.700 & 0.884 & 0.905 & 0.946 & 0.944 & 0.945 & 0.901 & 0.909 & 0.923 & 0.929 \\
			&300 & 1.000 & 0.946 & 0.943 & 0.762 & 0.777 & 0.901 & 0.922 & 0.950 & 0.943 & 0.941 & 0.910 & 0.915 & 0.942 & 0.939 \\ 
			&400 &  1.013 & 1.020 & 1.026 & 0.869 & 0.886 & 0.963 & 0.978 & 0.949 & 0.954 & 0.955 & 0.930 & 0.932 & 0.941 & 0.941 \\ 
			& &\multicolumn{14}{c}{WR under unbalanced design} \\
			\cmidrule(lr){3-16}
			&200&0.979 & 0.970 & 0.960 & 0.477 & 0.628 & 0.843 & 1.034 & 0.935 & 0.953 & 0.945 & 0.843 & 0.892 & 0.935 & 0.951 \\ 
			&300& 0.963 & 0.955 & 0.962 & 0.678 & 0.840 & 0.920 & 1.103 & 0.947 & 0.947 & 0.954 & 0.890 & 0.926 & 0.931 & 0.957 \\
			&400&0.951 & 0.953 & 0.967 & 0.779 & 0.957 & 0.925 & 1.112 & 0.932 & 0.942 & 0.944 & 0.911 & 0.945 & 0.941 & 0.953 \\ 
			& &\multicolumn{14}{c}{WD under unbalanced design} \\
			\cmidrule(lr){3-16}
			&200&1.006 & 1.036 & 1.024 & 0.508 & 0.664 & 0.882 & 1.073 & 0.945 & 0.955 & 0.958 & 0.839 & 0.883 & 0.930 & 0.960 \\ 
			&300&1.013 & 0.993 & 1.001 & 0.690 & 0.850 & 0.922 & 1.098 & 0.953 & 0.946 & 0.949 & 0.887 & 0.925 & 0.939 & 0.956 \\ 
			&400&0.966 & 0.978 & 0.993 & 0.784 & 0.951 & 0.929 & 1.101 & 0.940 & 0.951 & 0.949 & 0.909 & 0.933 & 0.941 & 0.961 \\ 
			\bottomrule
		\end{tabular}%
	\end{adjustbox}
\end{table}
\clearpage

\begin{table}
\footnotesize
	\begin{center}
		\caption{Empirical Type I error rate of estimated WR and WD from different estimators under balanced and unbalanced designs. UNADJ: the unadjusted estimator; IPW: inverse probability weighting; OW: overlap weighting; AIPW: augmented inverse probability weighting; AOW: augmented overlap weighting; AIPW-mis: augmented inverse probability weighting with a misspecified working outcome model; and AOW-mis: augmented overlap weighting with a misspecified working outcome model.}
		\label{tb:em_type_one}
		\begin{adjustbox}{width=1\textwidth}
			\begin{tabular}{llccrrrrrrr}
				\hline
				True model &Design & $n$ & Estimand  & UNADJ & IPW & OW & AIPW & AOW & AIPW-mis & AOW-mis   \\
				\hline
				Quadratic & Balanced &$200$ &WR &0.048 & 0.048 & 0.048 & 0.073 & 0.068 & 0.059 & 0.053 \\  
				&          &      &WD & 0.048 & 0.052 & 0.050 & 0.072 & 0.067 & 0.060 & 0.053 \\
				&          &$300$ &WR &0.052 & 0.060 & 0.059 & 0.069 & 0.066 & 0.059 & 0.051 \\ 
				&          &      &WD &0.055 & 0.060 & 0.060 & 0.069 & 0.066 & 0.059 & 0.053 \\  
				&          &$400$ &WR &0.040 & 0.048 & 0.051 & 0.041 & 0.041 & 0.046 & 0.045 \\  
				&          &      &WD &0.041 & 0.049 & 0.051 & 0.042 & 0.040 & 0.048 & 0.047 \\  
				\\
				&Unbalanced&$200$ &WR &0.056 & 0.065 & 0.067 & 0.121 & 0.069 & 0.088 & 0.049 \\ 
				&          &      &WD &0.063 & 0.067 & 0.070 & 0.123 & 0.067 & 0.091 & 0.048 \\   
				&          &$300$ &WR &0.060 & 0.055 & 0.057 & 0.081 & 0.043 & 0.077 & 0.043 \\  
				&          &      &WD &0.061 & 0.057 & 0.057 & 0.083 & 0.043 & 0.076 & 0.042 \\ 
				&          &$400$ &WR &0.042 & 0.049 & 0.055 & 0.071 & 0.040 & 0.064 & 0.035 \\ 
				&          &      &WD &0.046 & 0.050 & 0.056 & 0.075 & 0.038 & 0.065 & 0.036 \\  
				\\
				Interaction& Balanced &$200$ &WR &0.049 & 0.052 & 0.046 & 0.090 & 0.053 & 0.058 & 0.053 \\ 
				&          &      &WD &0.051 & 0.053 & 0.049 & 0.092 & 0.052 & 0.058 & 0.052 \\  
				&          &$300$ &WR &0.044 & 0.045 & 0.047 & 0.087 & 0.061 & 0.050 & 0.050 \\ 
				&          &      &WD &0.047 & 0.045 & 0.049 & 0.086 & 0.061 & 0.052 & 0.051 \\  
				&          &$400$ &WR &0.046 & 0.050 & 0.050 & 0.085 & 0.056 & 0.055 & 0.054 \\ 
				&          &      &WD &0.049 & 0.050 & 0.052 & 0.088 & 0.056 & 0.055 & 0.054 \\ 
				\\
				&Unbalanced&$200$ &WR &0.051 & 0.054 & 0.062 & 0.180 & 0.046 & 0.083 & 0.046 \\ 
				&          &      &WD &0.056 & 0.052 & 0.064 & 0.184 & 0.043 & 0.081 & 0.043 \\   
				&          &$300$ &WR &0.045 & 0.047 & 0.049 & 0.138 & 0.031 & 0.066 & 0.031 \\ 
				&          &      &WD &0.049 & 0.048 & 0.049 & 0.137 & 0.030 & 0.067 & 0.030 \\  
				&          &$400$ &WR &0.037 & 0.037 & 0.038 & 0.090 & 0.050 & 0.051 & 0.025 \\ 
				&          &      &WD &0.042 & 0.035 & 0.039 & 0.092 & 0.048 & 0.051 & 0.025 \\ 
				\hline 
			\end{tabular}
		\end{adjustbox}
	\end{center}
\end{table}

\begin{table}
\begin{center}
\caption{Data analyses results for primary and secondary ordinal outcomes in ORCHID trial by covariate-adjusted win statistics: point estimate for WR and WD (EST), standard error estimate calculated by corresponding closed-form variance estimation (SE), lower limit of the 95\% confidence intervals (LCL), upper limit of the 95\% confidence intervals (UCL), and proportion
	variance reduction compared to the unadjusted estimator (PVR). } 
		\label{tab:analysis1}
		\begin{adjustbox}{width=1\textwidth}
			\begin{tabular}{llccccccccccc}
				\hline
				&&\multicolumn{5}{c}{Win ratio} & &\multicolumn{5}{c}{Win difference (net benefit)} \\
				\cline{3-7}\cline{9-13}  Outcome & &UNADJ & IPW & OW & AIPW & AOW && UNADJ & IPW & OW & AIPW & AOW \\
				\hline
				Day 7 & EST &1.099 &1.062 &1.062 &1.066 &1.066 & &0.039 &0.025 &0.025 &0.026 &0.026\\
				& SE &0.143  &0.109 &0.109 &0.108 &0.108 & &0.054 &0.042 &0.042 &0.041 &0.042\\
				& LCL &0.818&0.849&0.849&0.855&0.855 &&-0.066&-0.058 & -0.058&-0.055&-0.055\\
				& UCL &1.379& 1.275&1.275&1.277&1.278&&0.144&0.107&0.107&0.108&0.108\\
				& PVR  & -   &41.9\%&41.9\% &43.3\%&42.8\%&&-&39.5\%&39.5\%&40.9\%&40.6\%\\
				\\
				Day 14 &EST &1.027 & 0.992 &0.995 &1.001&1.003&&0.010 &-0.003&-0.002&0.001 &0.001\\
				&SE  &0.142 & 0.115 &0.116 & 0.115 & 0.116 &&0.053 & 0.044& 0.044& 0.044 & 0.044\\
				&LCL &0.749 & 0.766 &0.768 & 0.776 & 0.776 &&-0.093&-0.089&-0.088&-0.085 &-0.085\\
				&UCL&1.306 & 1.218 &1.222 & 1.227 & 1.230 &&0.113 & 0.084& 0.085& 0.086 & 0.087\\
				&PVR &- &34.4\%&33.3\%&34.1\% &33.5\% && - &31.1\%&31.1\%&31.8\%&31.5\%\\
				\\
				Day 28 &EST &1.039 & 1.005 &1.008 & 0.998 & 1.001 &&0.013 & 0.002& 0.003&-0.001 & 0.000\\
				&SE  &0.159 & 0.132 &0.133 & 0.130 & 0.131 &&0.051 & 0.044& 0.044& 0.043 & 0.043\\
				&LCL &0.728 & 0.747 &0.749 & 0.743 & 0.744 &&-0.087& -0.084 &-0.083& -0.085 &-0.085\\
				&UCL &1.349& 1.264 &1.268 & 1.253 & 1.257 &&0.112 &  0.087 &0.088 & 0.084 &0.085\\
				&PVR &-&31.1\% &30.0\% &33.0\% &32.4\% && -&25.6\% &25.6\% &27.8\%&27.7\%\\
				\hline			
			\end{tabular}
		\end{adjustbox}
	\end{center}
\end{table}

\end{document}